\documentclass{ws-rv9x6}
\usepackage{subfigure}     
\usepackage{ws-rv-van}     
\makeindex

\begin{document}

\chapter[Quantum Metrology with Cold Atoms]{Quantum Metrology with Cold Atoms}\label{ra_ch1}

\author[J. Huang, S. Wu, H. Zhong, and C. Lee]{Jiahao Huang, Shuyuan Wu, Honghua Zhong, and Chaohong Lee\footnote{Corresponding author.}}

\address{State Key Laboratory of Optoelectronic Materials and Technologies,\\
School of Physics and Engineering, Sun Yat-Sen University, Guangzhou 510275, China\\
chleecn@gmail.com}

\begin{abstract}
Quantum metrology is the science that aims to achieve precision measurements by making use of quantum principles. Attribute to the well-developed techniques of manipulating and detecting cold atoms, cold atomic systems provide an excellent platform for implementing precision quantum metrology. In this chapter, we review the general procedures of quantum metrology and some experimental progresses in quantum metrology with cold atoms. Firstly, we give the general framework of quantum metrology and the calculation of quantum Fisher information, which is the core of quantum parameter estimation. Then, we introduce the quantum interferometry with single and multiparticle states. In particular, for some typical multiparticle states, we analyze their ultimate precision limits and show how quantum entanglement could enhance the measurement precision beyond the standard quantum limit. Further, we review some experimental progresses in quantum metrology with cold atomic systems.
\end{abstract}


\tableofcontents

\body

\section{Introduction}\label{sec1}

In recent years, the experimental techniques of manipulating cold atoms have been dramatically developed. Therefore, beyond exploring their quantum nature, it becomes possible to engineer cold atoms~\cite{RevModPhys.79.235, RevModPhys.82.1225, RevModPhys.80.885} for practical technology applications~\cite{RevModPhys.81.1051, PhilTransRSocLondA.361.1655}. Naturally, because of their robust quantum coherence and high controllability, cold atoms could be engineered to achieve high precision metrology at the level of quantum mechanics~\cite{Gross2010, Riedel2010, Anne2010, 12460641, L¨¹cke11112011}.

Quantum metrology~\cite{Giovannetti19112004, PhysRevLett.96.010401, Giovannetti2011} aims to yield high measurement precisions by taking advantage of the quantum principles. A central goal of quantum metrology is how to enhance measurement precision with quantum resources such as spin squeezing and multiparticle entanglement~\cite{Giovannetti19112004, PhysRevLett.96.010401, Giovannetti2011, NatPhoton.5.43, Lee2012, Gross2012, PhysRevA.46.R6797, Sorensen2001, PhysRevLett.97.150402, PhysRevLett.101.040403, PhysRevLett.102.100401, PhysRevLett.105.120501, Napolitano2011, Nagata04052007, Gross2011, PhysRevA.85.043604}. For a multiparticle system of cold atoms, it has been demonstrated that quantum entanglement and spin squeezing can be prepared by employing intrinsic inter-atom interactions or laser induced artificial inter-atom interactions~\cite{RevModPhys.73.307, Griffin1995, Pethick2002}. Up to now, cold atoms have been widely used for implementing precision metrology, such as interferometers~\cite{Roos2006, Yonezawa21092012, PhysRevA.61.041602}, gyroscopes~\cite{PhysRevA.81.043624}, quantum clocks~\cite{Swallows25022011, Martin09082013, PhysRevLett.92.230801}, magnetic field detectors~\cite{PhysRevA.82.022330, PhysRevA.87.043602, PhysRevLett.104.133601, arXiv:1303.1313} and micro-gravity sensors~\cite{PhysRevD.65.022002, PhysRevA.82.061602, PhysRevA.86.043630, PhysRevLett.110.093602}.

In this chapter, we review the recent progresses in quantum metrology with cold atoms. In Sec.~2, we present the general framework of quantum metrology. In particular, we give the general procedure of measurement in quantum mechanics and the fundamental theory of parameter estimation. In Sec.~3, we describe the basic principles of quantum interferometry with single-particle states, which includes the Ramsey interferometry and Mach-Zehnder interferometry. In Sec.~4, we show the basic principles of quantum interferometry with multiparticle states, such as, spin coherent states, spin squeezed states, NOON states, entangled coherent states and twin Fock states. In Sec.~5, we mention some key experimental progresses in quantum metrology with various cold atomic systems, such as, ultracold trapped ions, cold atomic ensembles and Bose-Einstein condensed atoms. In the last section, we briefly summarize this review and discuss some related problems.

\section{Quantum Metrology}

\subsection{Measurement in Quantum Mechanics}

At the level of physics, a measurement is a physical process which estimates the quantity of a particular observable (or a physical parameter)~\cite{Holevo1982}. It plays a key role in most natural sciences and practical technologies. To compare different measurements, one has to specify magnitude, units and uncertainty for a particular measurement. The science of measurement is called metrology.

The measurement process is governed by the laws of physics. Therefore, the measurement precision depends on both the performance imperfections and the fundamental limit imposed by the physical laws. The statistical fluctuations can be reduced by repeating the same measurement over times and averaging the results~\cite{PhysRevD.23.1693,Giovannetti2011}. According to the central limit theorem, for $N$ repetitions of the same measurement, the statistical fluctuation scales as $1/\sqrt{N}$ which is called the shot noise limit (SNL)~\cite{Giovannetti19112004}.

In quantum mechanics, a measurement process is the action that determines a particular observable (or a physical parameter) of a quantum system. Quantum metrology aims to make high-precision measurements with quantum resources such as entanglement and squeezing. It has been demonstrated that quantum metrology could give better precision than the same measurement performed in a classical framework. For an example, the measurement precision of a Mach-Zehnder interferometer of $N$ independent particles is limited by the standard quantum limit (SQL), which has the same scaling $1/\sqrt{N}$ as the SNL. However, the measurement precision of a Mach-Zehnder interferometer of $N$ entangled particles in the NOON state can reach the Heisenberg limit which has scaling $1/N$~\cite{PhysRevLett.85.2733,leea2002,PhysRevLett.105.180402}. Similar precision enhancements can also be obtained by other non-classical states such as spin squeezed states~\cite{Maccone2011}.

\subsection{General Procedure of Measurements}\label{2.2}

Usually, a general measurement process includes three steps~\cite{PhysRevLett.62.2377,Giovannetti2011,PhysRevLett.96.010401,
Escher2011}. First, prepare the probe into a desired initial state. Second, let the probe undergo a dynamical evolution dependent on the physical parameter to be measured. Third, read out the final state of the probe and estimate the physical parameter with the extracted information.

\begin{figure}[htb]
\centering
\includegraphics[width=4in]{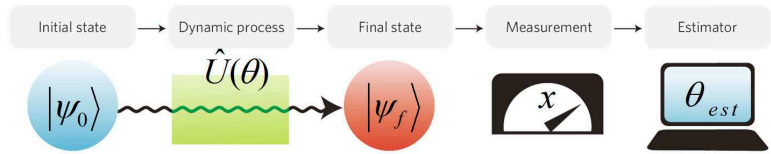}\caption{The general procedure for a measurement process in quantum mechanics. An initial state is sent through a physical channel and evolves into the final state under a parameter-dependent dynamical process. Then, the final state is read out and the unknown parameter is estimated. From Ref.~\citen{Escher2011}.}
\label{Fig1}
\end{figure}

The density matrix of the initial state $|\psi_{0}\rangle$ for the probe can be expressed as,
\begin{equation}\label{dm1}
\rho_{0}=|\psi_{0}\rangle \langle \psi_{0}|.
\end{equation}
The initial state is sent into a dynamical process dependent upon an unknown parameter $\theta$ to be measured. The initial state $|\psi_{0}\rangle$ evolves into the final state $|\psi(\theta)\rangle$ under the action of the parameter-dependent evolution operator $\hat{U}(\theta)$. If the evolution is unitary, the density matrix for the final state is given by,
\begin{equation}\label{dm2}
\rho(\theta)=|\psi(\theta)\rangle \langle \psi(\theta)|=\hat{U}(\theta) \rho_{0} \hat{U}^{\dagger}(\theta).
\end{equation}
Then, a measurement of a suitable observable $\hat{O}$ is made on the final state $|\psi(\theta)\rangle$. To successfully extract the unknown parameter $\theta$, the observable $\hat{O}$ should have $\theta$-dependent expectation values $\langle\hat{O}\rangle$.

If one has a prior knowledge of the evolution and the dependence of the observable expectation $\langle\hat{O}\rangle$ on the parameter $\theta$, the information of the parameter $\theta$ can be revealed according to the measurement results of the observable $\hat{O}$. According to the error propagation formula, the standard deviation of the parameter is given by
\begin{equation}\label{epf}
\Delta\theta=\frac{\Delta\hat{O}}{\left|\partial\langle \hat{O}\rangle /\partial\theta\right|},
\end{equation}
where the standard deviation of the observable is defined as
\begin{equation}\label{variance}
\Delta\hat{O}=\sqrt{\langle \hat{O}^2 \rangle-\langle \hat{O} \rangle^2}
\end{equation}
with
\begin{equation}\label{expectation1}
\langle \hat{O} \rangle = \langle \psi(\theta)| \hat{O} | \psi(\theta) \rangle,
\end{equation}
\begin{equation}\label{expectation2}
\langle \hat{O}^2 \rangle= \langle \psi(\theta)| \hat{O}^2 | \psi(\theta) \rangle.
\end{equation}
It clearly shows that the standard deviation of the parameter $\Delta \theta$ is dependent on $\theta$ itself. In addition, the minimum standard deviation of the parameter $\Delta \theta$ corresponds to the maximum slope of the expectation value with respect to the parameter $\theta$, $|\partial \langle \hat{O} \rangle /\partial\theta|$.

\subsection{Parameter Estimation}

Usually, measurement data are a set of random outputs from a particular effect dependent on the parameter to be estimated. The process of parameter estimation is using the information of a set of measurement data to estimate the value of the parameter. As several physical parameters can not be directly measured, one has to use indirect measurements, that is, inferring the parameter value from an estimator which is a function of the measurement data of a single observable or a set of observables.

There are two typical paradigms for implementing parameter estimation: global and local ones. In the global approach, an estimator which is independent on the value of the parameter is used to minimize a suitable cost function averaged all possible values of the parameter. In the local approach, an estimator which has minimum variance at a fixed value of the parameter is used to maximize the Fisher information. In most cases, since the optimization concerns a specific value of the parameter, the local approach is expected to provide a better ultimate bound on measurement precision. Below, we concentrate our discussion on the local approach~\cite{Escher2011,Helstrom1976,Kok2010}.

In a realistic estimation, one measures a suitable observable $X$ at first and then estimates the unknown parameter $\theta$ by an estimator function $\theta_{est}=T(X)$. Therefore, the deviation of an estimation is given by~\cite{PhysRevLett.72.3439, PhysRevLett.98.090401},
\begin{equation}\label{error}
\delta\theta\equiv\frac{\theta_{est}}{\left| d\left\langle
 \ensuremath{\theta_{est}}\right\rangle /d\theta \right|}-\theta,
\end{equation}
where $\theta_{est}$ and $\theta$ are the estimated and the actual values of the parameter, respectively. In the case of unbiased estimators, $\langle\theta_{est}\rangle=\theta$, it is just the difference between the estimated and the actual values of the parameter.

\subsubsection{Parameter estimation in classical statistics}

We consider the problem of estimating a single parameter with a set of measurement data from $v$ times of identical experiments (or one experiment of $v$ independent and identical probes). The measurement precision $\Delta \theta$ of an unknown parameter $\theta$ is limited by the Cram\'{e}r-Rao bound~\cite{Cramer1946,PhysRevLett.72.3439,PhysRevLett.69.3598}:
\begin{equation}
\Delta \theta =\sqrt{\left\langle (\delta\theta)^2 \right\rangle} \ge \frac{1}{\sqrt{vF(\theta)}}.
\end{equation}
Here, $F(\theta)$ is called the Fisher information, which is a measure of the ability to estimate a parameter $\theta$~\cite{Fisher1922,Fisher1925}. Obviously, the optimal measurement precision can be obtained by maximizing the Fisher information.

If the measured observable $X$ has $N$ different discrete values $(x_1, x_2, \cdots, x_N)$, the Fisher information $F(\theta)$ is expressed as
\begin{equation}\label{FI1}
F(\theta) \equiv \sum_{i=1}^{N} P\left(x_i|\theta\right) \left(\frac{\partial\ln[P(x_i|\theta)]} {\partial\theta}\right)^{2}
\end{equation}
with $P(x_i|\theta)$ denoting the conditional probability of the measurement data $x_i$ given the parameter $\theta$.

If the measured observable $X$ is a continuous variable $x$, the Fisher information $F(\theta)$ reads as
\begin{equation}\label{FI2}
F(\theta)\equiv\int dx p\left(x|\theta\right) \left(\frac{\partial\ln[p(x|\theta)]} {\partial\theta}\right)^{2}
\end{equation}
with $p(x|\theta)$ denoting the conditional probability density of the measurement data $x$ given the parameter $\theta$. Thus $p(x|\theta)dx$ represents the conditional probability of the measurement data between $x$ and $x+dx$ given the parameter $\theta$.

Now, we show why the Cram\'{e}r-Rao bound gives the minimum uncertainty of the estimated parameter~\cite{PhysRevLett.72.3439, PhysRevLett.69.3598, Matos2011,Kok2010}. We consider a set of measurement data $\left(X_{1}, X_{2}, ..., X_{v}\right)$ obtained from $v$ identical experiments (or one experiment of $v$ independent and identical probes). The parameter $\theta$ is estimated by a function constructed from the measurement data
\begin{equation}
\theta_{est}=T(x_{1}^{n_1},x_{2}^{n_2},\cdots,x_{v}^{n_v}),
\end{equation}
where $x_{j}^{n_j}$ is the $n_j$-th value of $X_j$ with $j=(1, 2,\cdots, v)$. Therefore, the averaged value of $\theta_{est}$ is given as
\begin{equation}
\left\langle \theta_{est}\right\rangle = \sum_{n_{1},n_{2},\cdots,n_{v}}  T\left(x_{1}^{n_1},x_{2}^{n_2},\cdots,x_{v}^{n_v}\right)  P\left(x_{1}^{n_1}|\theta\right)P\left(x_{2}^{n_2}|\theta\right) \cdots P\left(x_{v}^{n_v}|\theta\right).
\end{equation}
Obviously, the estimation function $\theta_{est}$ does not depend on the parameter $\theta$, but the average $\left\langle \theta_{est}\right\rangle$ depends on the parameter $\theta$.

Defining the deviation, $\delta\theta_{est} = \theta_{est} -\left\langle\theta_{est}\right\rangle$, one can easily find
\begin{equation}
\sum_{n_{1},n_{2},\cdots,n_{v}} P\left(x_1^{n_1}|\theta\right)P\left(x_2^{n_2}|\theta\right) \cdots P\left(x_v^{n_v}|\theta\right) \delta\theta_{est} = 0.
\end{equation}
According to the chain rule and $\partial\theta_{est} /\partial\theta = 0$, its derivative with respect to the parameter $\theta$ reads as,
\begin{eqnarray}
&&\sum_{n_{1},n_{2},\cdots,n_{v}}  \left\{ P\left(x_1^{n_1}|\theta\right)P\left(x_2^{n_2}|\theta\right) \cdots P\left(x_v^{n_v}|\theta\right)\right.\nonumber\\
&&\qquad\qquad\qquad \times \left[\sum_{i=1}^{v}\frac{1}{P(x_i^{n_i}|\theta)} \frac{\partial P(x_i^{n_i}|\theta)}{\partial\theta}\right] \left. \delta\theta_{est} \right\} -\frac{d\left\langle\theta_{est}\right\rangle} {d\theta}=0.
\end{eqnarray}
Therefore,
\begin{eqnarray}\label{eq1}
\frac{d\left\langle \theta_{est}\right\rangle }{d\theta} =&& \sum_{n_{1},n_{2},\cdots,n_{v}} \left\{ P\left(x_1^{n_1}|\theta\right)P\left(x_2^{n_2}|\theta\right) \cdots P\left(x_v^{n_v}|\theta\right)\right.\nonumber\\
&&\qquad\qquad\qquad\qquad \times \left(\sum_{i=1}^{v} \frac{\partial\ln\left[P(x_i^{n_i}|\theta)\right]} {\partial\theta}\right) \left.\delta\theta_{est}\right\}.
\end{eqnarray}
By applying the inequality $\left\langle AB\right\rangle^2 \leq \left\langle A^2\right\rangle \left\langle B^2\right\rangle$ to the right-hand side of the above equation, one can obtain
\begin{eqnarray}\label{eq2}
&& \left\{ \sum_{n_{1},n_{2},\cdots,n_{v}} P\left(x_1^{n_1}|\theta\right)P\left(x_2^{n_2}|\theta\right) \cdots P\left(x_v^{n_v}|\theta\right) \left(\sum_{i=1}^{v} \frac{\partial\ln\left[P(x_i^{n_i}|\theta)\right]} {\partial\theta}\right)^{2} \right\}\times \nonumber\\
&& \left\{\sum_{n_{1},n_{2},\cdots,n_{v}} P\left(x_1^{n_1}|\theta\right)P\left(x_2^{n_2}|\theta\right) \cdots P\left(x_v^{n_v}|\theta\right) \left(\delta\theta_{est}\right)^{2}\right\}
\geq \left(\frac{d\left\langle \theta_{est}\right\rangle }{d\theta}\right)^{2}.
\end{eqnarray}
Since
\begin{equation}
\sum_{{n_i}}P(x_i^{n_i}|\theta) \frac{\partial\ln[P(x_i^{n_i}|\theta)]} {\partial\theta} =\sum_{{n_i}}\frac{\partial P(x_i^{n_i}|\theta)}{\partial\theta} =\frac{d}{d\theta}\sum_{{n_i}}P(x_i^{n_i}|\theta)=0,
\end{equation}
all cross terms in the right-hand side of Eq.~\eqref{eq2} vanish, and the square of the sum reduces to
\begin{eqnarray}
&&\left\{ \sum_{n_{1},n_{2},\cdots,n_{v}} P\left(x_1^{n_1}|\theta\right)P\left(x_2^{n_2}|\theta\right) \cdots P\left(x_v^{n_v}|\theta\right) \sum_{i=1}^{v} \left(\frac{\partial\ln\left[P(x_i^{n_i}|\theta)\right]} {\partial\theta}\right)^{2}\right\} \times \nonumber\\
&& \left\{\sum_{n_{1},n_{2},\cdots,n_{v}} P\left(x_1^{n_1}|\theta\right)P\left(x_2^{n_2}|\theta\right) \cdots P\left(x_v^{n_v}|\theta\right) \left(\delta\theta_{est}\right)^{2}\right\} \geq
\left(\frac{d\left\langle \theta_{est}\right\rangle }{d\theta}\right)^{2}.
\end{eqnarray}
Denoting the averaged square deviation of $\theta_{est}$ as
\begin{equation}
\left\langle \left(\delta\theta_{est}\right)^{2}\right\rangle =\sum_{n_{1},n_{2},\cdots,n_{v}}^{} P\left(x_1^{n_1}|\theta\right)P\left(x_2^{n_2}|\theta\right) \cdots P\left(x_v^{n_v}|\theta\right) \left(\delta\theta_{est}\right)^{2},
\end{equation}
one can obtain
\begin{eqnarray}
&& \left\{\sum_{n_{1},n_{2},\cdots,n_{v}} P\left(x_1^{n_1}|\theta\right)P\left(x_2^{n_2}|\theta\right) \cdots P\left(x_v^{n_v}|\theta\right) \sum_{i=1}^{v} \left(\frac{\partial\ln\left[P(x_i^{n_i}|\theta)\right]} {\partial\theta}\right)^{2}\right\}\nonumber\\
&& \times \left\langle \left(\delta\theta_{est}\right)^{2}\right\rangle \geq
\left(\frac{d\left\langle \theta_{est}\right\rangle }{d\theta}\right)^{2}.
\end{eqnarray}
As the measurement data $\left(X_{1},X_{2},...,X_{v}\right)$ are independent and the sum for each $X_{j}$ has the same form of the Fisher information~\eqref{FI1}, the above inequality becomes
\begin{equation}
vF(\theta) \left\langle \left(\delta\theta_{est}\right)^{2}\right\rangle \geqslant \left(\frac{d\left\langle \theta_{est}\right\rangle }{d\theta}\right)^{2}.
\end{equation}
It can also be written in the form of
\begin{equation}\label{ineq1}
\frac{\left\langle \left(\delta\theta_{est}\right)^{2}\right\rangle} {\left(d\left\langle \theta_{est}\right\rangle /d\theta\right)^{2}} \geqslant \frac{1}{vF(\theta)}.
\end{equation}

To show how the estimated parameter is close to the actual one, the uncertainty given in Eq.~\eqref{error} should be calculated. The average of the square of Eq.~\eqref{error} is given as
\begin{equation}\label{D1}
\left\langle \left(\delta\theta\right)^{2}\right\rangle =\frac{\left\langle \delta\theta_{est}^{2}\right\rangle} {\left(d\left\langle \theta_{est}\right\rangle /d\theta\right)^{2}} +\left\langle \delta\theta\right\rangle ^{2}.
\end{equation}
Substituting Eq.~\eqref{D1} into Eq.~\eqref{ineq1}, the inequality reads as
\begin{equation}
\left\langle \left(\delta\theta\right)^{2}\right\rangle \geqslant\frac{1}{vF(\theta)} +\left\langle \delta\theta\right\rangle ^{2}.
\end{equation}
In general, we regard the expectation of the unknown estimated parameter over many times as the actual one, i.e., $\langle \theta_{est}\rangle =\theta$. Therefore, the derivative $d\langle \theta_{est}\rangle / d\theta=1$, $\left\langle \delta\theta\right\rangle ^{2}=0$, and $\left\langle (\delta\theta)^{2}\right\rangle$ reduces to the variance
$\left(\Delta\theta\right)^{2}$. That is, the above inequality becomes
\begin{equation}
\left(\Delta\theta\right)^2 = \left\langle \left(\delta\theta\right)^{2}\right\rangle \geqslant \frac{1}{vF(\theta)},
\end{equation}
whose square root,
\begin{equation}\label{CR1}
\Delta\theta =\sqrt{\left\langle \left(\delta\theta\right)^{2} \right\rangle} \geqslant \frac{1}{\sqrt{vF(\theta)}},
\end{equation}
gives the so-called Cram\'{e}r-Rao bound.

If the measured observable $X$ is a continuous variable $x$, the deviation of the Cram\'{e}r-Rao bound can be obtained by replacing the sum $\sum_{n_1,n_2,\cdots,n_v}$ with the integral $\int dx_{1}dx_{2} \cdots dx_{v}$. The form of the Cram\'{e}r-Rao bound is as same as the formula~\eqref{CR1} and the corresponding Fisher information is given by the formula~\eqref{FI2}.

\subsubsection{Parameter estimation in quantum mechanics}

In quantum mechanics, a generalized measurement can be described by a set of Hermitian operators ${\hat{E}(X)}$, which are Positive-Operator Valued Measures (POVM)~\cite{Nielsen2001}. Here, $X$ is the measured observable. If $X$ has $N$ different discrete values $(x_1, x_2, \cdots, x_N)$, the operators $\hat{E}(x_n)$ satisfy,
\begin{equation}
\hat{E}(x_n) \ge 0, \qquad \sum_{n=1}^{N} \hat{E}(x_n)=\bf{1},
\end{equation}
where $\bf{1}$ is the identity operator. If $X$ has continuous values $x$, the operators $\hat{E}(x)$ satisfy,
\begin{equation}
\hat{E}(x)\ge 0, \qquad \int dx \hat{E}(x)=\bf{1}.
\end{equation}
The above relations ensure non-negative probabilities and unitary total probability.

The probability of getting a particular measurement data $x_n$ on a state $\left|\psi\right\rangle$ is given by,
\begin{equation}\label{p1}
P(x_n) = \left\langle \psi |\hat{E}(x_n)|\psi\right\rangle = Tr[\rho \hat{E}(x_n)],
\end{equation}
where $\rho=\left|\psi\rangle \langle \psi\right|$ is the density matrix. Therefore, by making a measurement on the final state, one can obtain the conditional probability of the measurement data $x_n$ given the parameter $\theta$,
\begin{equation}\label{probability1}
P(x_n|\theta)=\left\langle \psi(\theta)| \hat{E}(x_n) |\psi(\theta)\right\rangle =Tr[\rho(\theta)\hat{E}(x_n)].
\end{equation}
If $X$ has continuous values $x$, the corresponding conditional probability density of the measurement data $x$ given the parameter $\theta$ reads as,
\begin{equation}\label{probability2}
p(x|\theta)=\left\langle \psi(\theta)| \hat{E}(x) |\psi(\theta)\right\rangle =Tr[\rho(\theta)\hat{E}(x)].
\end{equation}

According to the definition of the Fisher information~\eqref{FI1}, one can find that $F(\theta)$ is a function of all conditional probabilities $P(x_n|\theta)$ which depend on the final state $\left|\psi(\theta)\right\rangle$ and the POVM $\hat{E}(x_n)$. Therefore, to optimize the Fisher information, we need to construct a suitable state and measure a suitable observable. Further, for a given final state, the Fisher information can be maximized by trying different measurement strategies. The maximum of the Fisher information through out all possible quantum measurement strategies is called the quantum Fisher information~\cite{Helstrom1976, PhysRevLett.72.3439,AnnPhys.247.135},
\begin{equation}
F_{Q}(\theta) \equiv \max_{\{\hat{E}(x_n)\}} F\left[\theta; \{\hat{E}(x_n)\}\right],
\end{equation}
where $\{\hat{E}(x_n)\}=\left(\hat{E}(x_1), \hat{E}(x_2), \cdots, \hat{E}(x_N)\right)$. The corresponding Cram\'{e}r-Rao bound is called the quantum Cram\'{e}r-Rao bound,
\begin{equation}
\Delta\theta = \sqrt{\left\langle (\delta\theta)^{2}\right\rangle }\geqslant\frac{1}{\sqrt{vF_{Q}(\theta)}}.
\end{equation}

We now show how to derive the quantum Fisher information and the quantum Cram\'{e}r-Rao bound for a pure state under a unitary evolution~\cite{Matos2011}. Here, we consider a probe in the initial state $\left|\psi_{0}\right\rangle$ undergoing an evolution described by the operator $\hat{U}(\theta)=\exp\left(-i\hat{H}\theta\right)$ with the generator $\hat{H}$. Inserting the density matrix~\eqref{dm2} into the conditional probality~\eqref{probability1}, we obtain
\begin{eqnarray}\label{p2}
\frac{\partial P(x_n|\theta)}{\partial\theta} & = & \left[\frac{d}{d\theta}\langle\psi(\theta)|\right]\hat{E}(x_n) |\psi(\theta)\rangle +\langle\psi(\theta)|\hat{E}(x_n) \left[\frac{d}{d\theta}|\psi(\theta)\rangle\right] \nonumber\\
 & = & i\langle\psi(\theta)|\hat{H}\hat{E}(x_n)|\psi(\theta)\rangle -i\langle\psi(\theta)|\hat{E}(x_n)\hat{H}|\psi(\theta)\rangle \nonumber \\
 & = & i\langle\psi(\theta)|\left[\hat{H},\hat{E}(x_n)\right] |\psi(\theta)\rangle \nonumber \\
 & = & - 2\textrm{Im}\left[\langle\psi(\theta)|\hat{H}\hat{E}(x_n) |\psi(\theta)\rangle\right].
\end{eqnarray}
Introducing an arbitrary real function $G(\theta)$ into Eq.~\eqref{p2}, we have
\begin{equation}
\frac{\partial P(x_n|\theta)}{\partial\theta}=-2\textrm{Im}\left\{ \langle\psi(\theta)|\left[\hat{H}-G(\theta)\right] \hat{E}(x_n)|\psi(\theta)\rangle\right\},
\end{equation}
and so that
\begin{eqnarray}
&& \left(\frac{\partial P(x_n|\theta)}{\partial\theta}\right)^{2}
= 4\left\{\textrm{Im}\left\{ \langle\psi(\theta)|\left[\hat{H}-G(\theta)\right]\hat{E}(x_n) |\psi(\theta)\rangle\right\}\right\}^2 \nonumber\\
&& \leqslant 4|\langle\psi(\theta)|\left[\hat{H}-G(\theta)\right]\hat{E}(x_n) |\psi(\theta)\rangle|^{2} \\
&& \leqslant 4\langle\psi(\theta)|\hat{E}(x_n)|\psi(\theta)\rangle \langle\psi(\theta)|\left[\hat{H}-G(\theta)\right] \hat{E}(x_n)\left[\hat{H}-G(\theta)\right] \left|\psi(\theta)\right\rangle \nonumber\\
&& =  4P(x_n|\theta)\langle\psi(\theta)|\left[\hat{H}-G(\theta)\right] \hat{E}(x_n)\left[\hat{H}-G(\theta)\right] |\psi(\theta)\rangle.\nonumber
\end{eqnarray}
Therefore, one can obtain
\begin{eqnarray}
F(\theta) & = & \sum_{n} \frac{1}{P(x_n|\theta)} \left[\frac{\partial P(x_n|\theta)}{\partial\theta}\right]^{2} \\
 & \leqslant & 4\sum_{n} \langle\psi(\theta)|\left[\hat{H}-G(\theta)\right] \hat{E}(x_n)\left[\hat{H}-G(\theta)\right] |\psi(\theta)\rangle \nonumber \\
 & = & 4\langle\psi(\theta)|\left[\hat{H}-G(\theta)\right]^{2} |\psi(\theta)\rangle.\nonumber
\end{eqnarray}
If $X$ has continues values, the corresponding Fisher information reads as
\begin{eqnarray}
F(\theta) & = & \int dx \frac{1}{p(x|\theta)} \left[\frac{\partial p(x|\theta)}{\partial\theta}\right]^{2} \\
 & \leqslant & 4\int dx \langle\psi(\theta)|\left[\hat{H}-G(\theta)\right] \hat{E}(x)\left[\hat{H}-G(\theta)\right] |\psi(\theta)\rangle \nonumber \\
 & = & 4\langle\psi(\theta)|\left[\hat{H}-G(\theta)\right]^{2} |\psi(\theta)\rangle.\nonumber
\end{eqnarray}

If we choose $G(\theta)=\langle\psi_{0}|\hat{H}|\psi_{0}\rangle$, due to $\langle\hat{H}\rangle=\langle\psi_{0}|\hat{H}|\psi_{0}\rangle =\langle\psi(\theta)|\hat{H}|\psi(\theta)\rangle$, the Fisher information attains its minimum and the above inequality reads as
\begin{equation}
F(\theta)\leqslant4 \langle\psi_{0}|(\Delta\hat{H})^{2}|\psi_{0}\rangle
\end{equation}
with $\Delta\hat{H}=\hat{H}-\langle\hat{H}\rangle$. Therefore, the quantum Fisher information for a pure state can be defined as
\begin{equation}\label{FI3}
F_{Q}(\theta) =4\langle\psi_{0}|(\Delta\hat{H})^{2}|\psi_{0}\rangle,
\end{equation}
which is a function of the generator $\hat{H}$ and the initial state $\left|\psi_{0}\right\rangle$.

By using $\left|\psi(\theta)\right\rangle =\hat{U}\left|\psi_{0}\right\rangle =\exp(-i\hat{H}\theta)\left|\psi_{0}\right\rangle$ and $\left|\psi'(\theta)\right\rangle =\frac{d}{d\theta}\left|\psi(\theta)\right\rangle =(-i\hat{H})\left|\psi(\theta)\right\rangle$, the quantum Fisher information can be expressed as
\begin{equation}\label{FI4}
F_{Q}(\theta)=4\left[\langle\psi'(\theta)|\psi'(\theta)\rangle -\left|\langle\psi'(\theta)|\psi(\theta)\rangle\right|^{2}\right],
\end{equation}
which is a function of the final state $\left|\psi(\theta)\right\rangle$ and its derivative with respect to the parameter $\theta$.

The quantum Fisher information provides a powerful tool for parameter estimation only dependent on the state of the system but not on the procedure of measurement~\cite{PhysRevA.83.061802, PhysRevLett.102.040403, PhysRevA.80.013825, EPL.93.64002}. As long as the initial state and the final state of the probe after the parameter-dependent evolution are known, one can immediately predict the minimum uncertainty of the parameter to be estimated. Generally speaking, the quantum Cram\'{e}r-Rao bound is the ultimate bound on the parameter uncertainty, and the parameter uncertainty in realistic measurements may be larger.

\section{Quantum Interferometry with Single-Particle States}

Interferometry is an important and most used method for implementing measurements~\cite{RevModPhys.84.777, PhysRevA.79.033822, PhysRevLett.76.4295, PhysRevLett.107.207002}. Interferometry via quantum states includes three key steps: (i) splitting the initial state into two modes, (ii) undergoing a period of free evolution and (iii) recombining two modes for readout. There are two typical types of interferometry. One is the Mach-Zehnder interferometry, which has extensive applications in phase shift measurements~\cite{PhysRevLett.99.223602,PhysRevLett.99.070801}. The other is the Ramsey interferometry, which has been widely used in atomic-molecular experiments for precision spectroscopy and measurement~\cite{PhysRev.78.695,PhysRev.76.996,PhysRev.55.526}. In this section, we briefly introduce these two kinds of interferometry with single-particle states.

It is well known that a two-mode (or two-level) quantum particle can be regarded as a spin-$\frac{1}{2}$ particle, which can be described by the three Pauli matrices $\hat{\sigma}_{x}$, $\hat{\sigma}_{y}$ and $\hat{\sigma}_{z}$~\cite{Foot2005,PhysRevA.84.043628}. The two eigenstates $\left|\uparrow\right\rangle$ and $\left|\downarrow\right\rangle$ obey $\frac{\sigma_{z}}{2}\left|\uparrow\right\rangle =+\frac{1}{2}\left|\uparrow\right\rangle$ and $\frac{\sigma_{z}}{2}\left|\downarrow\right\rangle =-\frac{1}{2}\left|\downarrow\right\rangle$ and an arbitrary pure state can be written as $|\theta,\varphi\rangle =e^{i\gamma}\left(\sin\frac{\theta}{2}\left|\uparrow\right\rangle +\cos\frac{\theta}{2}e^{i\varphi}\left|\downarrow\right\rangle\right)$ with the common phase $\gamma$. The factor $e^{i\gamma}$ has no observable effects, thus the pure states $|\theta,\varphi\rangle$ with different values of $\gamma$ are represented by the same classical spin $(S_{x},S_{y},S_{z})=\frac{1}{2} (\sin\theta\cos\varphi,\sin\theta\sin\varphi,\cos\theta)$ in the Bloch sphere. Where, the longitudinal component $S_{z}=\cos\theta=\frac{1}{2} \left(\cos^2\frac{\theta}{2}-\sin^2\frac{\theta}{2}\right)$ stands for the half population difference between the two eigenstates, and the transverse components $(S_{x}, S_{y})$ stand for the quantum coherence between the two modes. This means that the polar angle $\theta$ reflects the polarization information, while the azimuthal angle $\varphi$ corresponds to coherence.

\subsection{Mach-Zehnder interferometry}

\begin{figure}[htb]
\centering
\includegraphics[width=4in]{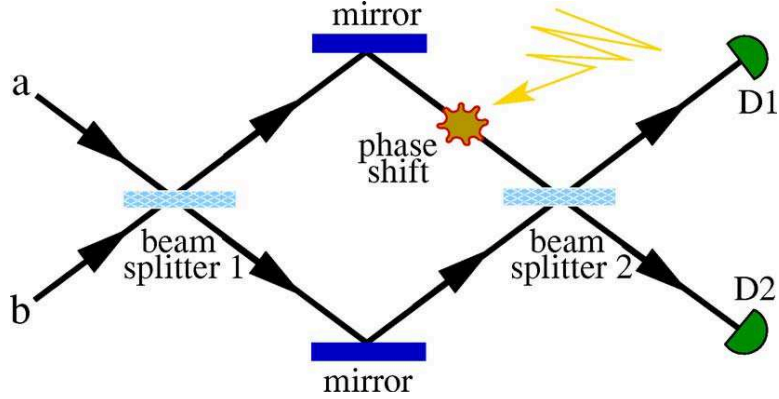}\caption{Schematic diagram of a Mach-Zehnder interferometer. Single atoms/photons enter the input ports, combine in the first beam splitter, evolve in the two paths, recombine via the second beam splitter and are finally detected in D1 and D2. The phase shift is inferred from the number of atoms or photons measured in each output port. From Ref.~\citen{PhysRevA.73.011801}.}
\label{fig_MZ}
\end{figure}

A conventional Mach-Zehnder interferometer is composed of two beam splitters and two propagation paths~\cite{PhysRevA.63.053804,PhysRevLett.99.223602,PhysRevLett.110.163604}. A collimated beam of single particles is divided into two parts by a 50:50 beam splitter. Then the two parts pass through two different spatial paths and accumulate a relative phase shift between the two parts. At last, the two parts are recombined for interference via another 50:50 beam splitter. The phase difference can be extracted from the interference fringe~\cite{PhysRevLett.89.247901}. The schematic diagram is shown in Fig.~\ref{fig_MZ}.

Suppose an atom incidents in the input port $a$, that is, the initial state of the atom is prepared in mode $|a\rangle$. The first beam splitter transforms the input state into an equal superposition state of the two involved modes $|a\rangle$ and $|b\rangle$,
\begin{equation}
|\psi_{in}\rangle =\hat{T} |a \rangle = \frac{1}{\sqrt{2}}(|a \rangle +|b \rangle),
\end{equation}
with the transformation matrix,
\begin{equation}
\hat{T} = \frac{1}{\sqrt{2}}
          \left(
            \begin{array}{cc}
              1 & 1 \\
              1 & -1 \\
            \end{array}
          \right).
\end{equation}
Then the two modes propagate along different paths and accumulate a relative phase shift $\varphi$. That is, before entering into the second beam splitter, the state reads as
\begin{equation}
|\psi_{out}\rangle=\frac{1}{\sqrt{2}}(|a \rangle + e^{i\varphi}|b \rangle).
\end{equation}
The second beam splitter recombines the two paths and the state is transformed into,
\begin{equation}
|\psi_{f}\rangle=\hat{T}\psi_{out} =\frac{1}{\sqrt{2}}[(1+e^{i\varphi})|a \rangle + (1-e^{i\varphi})|b \rangle].
\end{equation}
At last, the output state is detected by $D1$ and $D2$, which give $p(a|\varphi)=\cos^2(\varphi/2)$ for the probability of the atom in $|a\rangle$ and $p(b|\varphi)=\sin^2(\varphi/2)$ for the probability of the atom in $|b\rangle$.

From Eq.~\eqref{FI2}, the Fisher information for the above single-atom Mach-Zehnder interferometry can be obtained,
\begin{eqnarray}
F(\varphi)& = & \frac{1}{p(a|\varphi)}[\frac{\partial p(a|\varphi)}{\partial \varphi}]^2
+ \frac{1}{p(b|\varphi)}[\frac{\partial p(b|\varphi)}{\partial \varphi}]^2, \\
& = & \sin^2 \varphi + \cos^2 \varphi =1. \nonumber
\end{eqnarray}
Therefore, the minimal uncertainty of the relative phase is given as,
\begin{equation}\label{FI_R}
\delta \varphi = \frac{1}{\sqrt{F(\varphi)}} = 1.
\end{equation}
Repeating the experiment $N$ times, the uncertainty of the relative phase reads as,
\begin{equation}
\delta \varphi = \frac{1}{\sqrt{N F(\varphi)}} = \frac{1}{\sqrt{N}},
\end{equation}
which attains the so-called standard quantum limit (SQL).

\subsection{Ramsey interferometry}

The conventional Ramsey interferometry consists of two $\frac{\pi}{2}$ pulses and a free evolution process. In comparison with Mach-Zehnder interferometry, the two $\frac{\pi}{2}$ pulses act as the two beam splitters and the free evolution accumulates the relative phase between the two involved levels~\cite{Serge2013,Kleppner2013,Ramsey1980,PhysRevA.54.R4649}.

We consider a two-level atom, which is initially prepared in its ground state $\left|\downarrow\right\rangle$. Without loss of generality, we assume that the eigen-energies for the ground state $\left|\downarrow\right\rangle$ and the excited state $\left|\uparrow\right\rangle$ are $-\frac{\omega_0}{2}$ and $+\frac{\omega_0}{2}$, respectively. The first $\frac{\pi}{2}$ pulse is applied and the state becomes
\begin{equation}
|\psi_{in}\rangle=\frac{1}{\sqrt{2}} \left(\left|\downarrow\right\rangle +\left|\uparrow\right\rangle\right),
\end{equation}
which is an equal-probability superposition of the ground state $\left|\downarrow\right\rangle$ and the excited state $\left|\uparrow\right\rangle$.
Then, the system undergoes a free evolution. In which, the ground state $\left|\downarrow\right\rangle$ accumulates a negative phase $-\frac{\varphi}{2}$, while the excited state $\left|\uparrow\right\rangle$ accumulates a positive phase $+\frac{\varphi}{2}$. Therefore, the state after the free evolution reads as,
\begin{equation}
|\psi_{out}\rangle=\frac{1}{\sqrt{2}} (e^{-i\varphi/2}\left|\downarrow\right\rangle + e^{+i\varphi/2}\left|\uparrow\right\rangle).
\end{equation}
At last, the second $\frac{\pi}{2}$ pulse is applied and the final state is measured. The probability of the atom in the ground state $\left|\downarrow\right\rangle$ reads as,
\begin{equation}
p(\downarrow|\varphi) = \frac{1+\cos\varphi}{2} = \cos^2 \frac{\varphi}{2},
\end{equation}
and the probability of the atom in the excited state $\left|\uparrow\right\rangle$ reads as,
\begin{equation}
p(\uparrow|\varphi) = 1-p(\downarrow|\varphi)= \frac{1-\cos\varphi}{2} = \sin^2 \frac{\varphi}{2}.
\end{equation}

The relative phase $\varphi$ can be estimated from the probability $p(\downarrow|\varphi)$ or $p(\uparrow|\varphi)$. From Eq.~\eqref{FI2}, one can obtain the Fisher information for the above single-atom Ramsey interferometry,
\begin{eqnarray}
F(\varphi)& = & \frac{1}{p(\downarrow|\varphi)}[\frac{\partial p(\downarrow|\varphi)}{\partial \varphi}]^2
+ \frac{1}{p(\uparrow|\varphi)}[\frac{\partial p(\uparrow|\varphi)}{\partial \varphi}]^2, \\
& = & \sin^2 \varphi + \cos^2 \varphi =1. \nonumber
\end{eqnarray}
Therefore, the minimal uncertainty of the relative phase is given by,
\begin{equation}\label{FI_R}
\delta \varphi = \frac{1}{\sqrt{F(\varphi)}} = 1.
\end{equation}
Repeating the experiment $N$ times, the uncertainty of the relative phase is limited by the standard quantum limit,
\begin{equation}
\delta \varphi = \frac{1}{\sqrt{N F(\varphi)}} = \frac{1}{\sqrt{N}}.
\end{equation}

In the Bloch sphere, the initial state $\left|\downarrow\right\rangle$ is denoted by a spin vector pointing from the origin to the south pole. The first $\frac{\pi}{2}$ pulse rotates the state an angle $\frac{\pi}{2}$ around the $y$-axis. The free evolution rotates the state an angle $\varphi$ around the $z$-axis. The second $\frac{\pi}{2}$ pulse rotates the state an angle $\frac{\pi}{2}$ around the $y$-axis. Lastly, the angle between the spin vector for the final state and the $z$-axis is just the angle $\varphi$.

\begin{figure}[htb]
\centering
\includegraphics[width=4in]{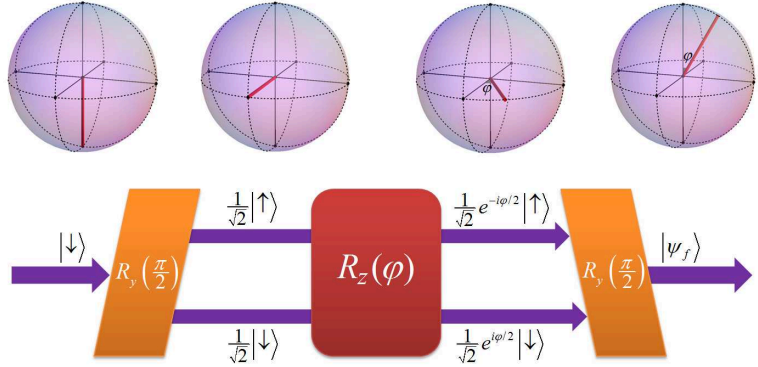}\caption{Schematic diagram for a single-atom Ramsey interferometer. The two $\frac{\pi}{2}$ pulses and the free evolution are specific rotations in the geometrical representation via the Bloch sphere.}
\label{fig_Ramsey}
\end{figure}

The state evolution from the initial state to the final state can be described by,
\begin{equation}
|\psi_{f}\rangle=\hat{U}|\psi_{0}\rangle,
\end{equation}
with the propagation operator
\begin{equation}\label{PO}
\hat{U} = \exp(-i\frac{\pi}{2}\hat{S}_{y}) \exp(-i\varphi\hat{S}_{z})\exp(-i\frac{\pi}{2}\hat{S}_{y}).
\end{equation}
Here, the spin operators are defined as $\hat{S}_{x,y,z}=\frac{1}{2}\hat{\sigma}_{x,y,z}$. Thus the expectation value for the final state reads as,
\begin{equation}\label{Sz}
\langle \hat{S}_{z} \rangle _{f} = \langle \psi_{f}|\hat{S}_{z}|\psi_{f}\rangle = \langle\psi_{0}|\hat{U}^{\dagger}\hat{S}_{z} \hat{U}|\psi_{0}\rangle.
\end{equation}
Substituting Eq.~\eqref{PO} into Eq.~\eqref{Sz}, we have
\begin{equation}
\langle \hat{S}_{z} \rangle_{f} = -\cos \varphi \langle \hat{S}_{z} \rangle_{0} + \sin \varphi \langle \hat{S}_{y} \rangle_{0}
\end{equation}
and
\begin{eqnarray}
 \left(\Delta \hat{S}_{z}\right)^2_{f}
& = & \left(\Delta \hat{S}_{z}\right)^2_{0} \cos^2 \varphi +\left(\Delta \hat{S}_{y}\right)^2_{0} \sin^2 \varphi\\
& - & \sin \varphi \cos \varphi \left\langle \hat{S}_{z}\hat{S}_{y}+\hat{S}_{y}\hat{S}_{z} \right\rangle \nonumber
\end{eqnarray}
As the initial state $|\psi_{0}\rangle=\left|\downarrow\right\rangle_{0}$, we have
\begin{equation}
\langle\hat{S}_{z}\rangle_{0}=\frac{1}{2},
\langle\hat{S}_{z}^2\rangle_{0}=\langle\hat{S}_{y}^2\rangle_{0}=\langle\hat{S}_{x}^2\rangle_{0}=\frac{1}{4},
\langle\hat{S}_{x}\rangle_{0}=\langle\hat{S}_{y}\rangle_{0}=0. \nonumber
\end{equation}
Thus one can immediately obtain
\begin{equation}
\left(\Delta \hat{S}_{z}\right)_{f} = \frac{1}{2} \sin \varphi
\end{equation}
According to Eq.~\eqref{epf}, the standard deviation of $\varphi$ is given by,
\begin{equation}
\Delta \varphi = \frac{\left(\Delta \hat{S}_{z}\right)_{f} }{\left|\partial\langle \hat{S}_{z} \rangle _{f} /\partial\varphi\right|} = 1,
\end{equation}
which agrees with the minimal uncertainty given by the Fisher information Eq.~\eqref{FI_R}. Repeating the measurement $N$ times, the uncertainty can reach the standard quantum limit $\Delta \varphi \sim \frac{1}{\sqrt{N}}$.

\section{Quantum Interferometry with Multiparticle States}

In this section, we discuss the quantum interferometry with multiparticle states. There are many different types of multiparticle states that have been used to implement quantum interferometry~\cite{PhysRevLett.71.1355, PhysRevLett.79.3865, PhysRevLett.85.5098, PhysRevLett.91.010402, PhysRevLett.104.103602, PhysRevLett.105.013603, Kaushik2011, Ben-Aryeh:12, PhysRevA.50.4228, PhysRevA.61.043811, PhysRevA.68.023810, PhysRevA.68.033821, PhysRevA.78.063828, PhysRevA.82.013831, PhysRevA.87.023825, NewJPhys.11.033002}. Here, we concentrate our discussions on coherent spin states, spin squeezed states, NOON states, entangled coherent states and twin Fock states.

\subsection{Coherent Spin States}

For an ensemble of $N$ two-level particles, which can be regraded as $N$ identical spin-$\frac{1}{2}$ particles, one can mathematically describe the system by a collective spin of length $J=\frac{N}{2}$~\cite{Gross2012,PhysRevA.6.2211}. Such a spin-J system is characterized by three collective spin operators $\hat{J}_{x}$, $\hat{J}_{y}$ and $\hat{J}_{z}$ which are defined as the sum of spin operators of spin-$\frac{1}{2}$ spin operators $\hat{S}_{x}$, $\hat{S}_{y}$ and $\hat{S}_{z}$,
\begin{equation}
\hat{J_{i}}={\displaystyle \sum_{l=1}^{N}\hat{S}_{i}^{(l)}},(i=x,y,z).
\end{equation}
By using the Schwinger representation~\cite{Sakurai2005}, the collective spin operators can be written in the form of
\begin{equation}\label{CSO1}
\hat{J_{x}}=\frac{1}{2}\left(\hat{a}^{\dagger}\hat{b} +\hat{b}^{\dagger}\hat{a}\right),
\end{equation}
\begin{equation}\label{CSO2}
\hat{J_{y}}=\frac{i}{2}\left(\hat{a}^{\dagger}\hat{b} -\hat{b}^{\dagger}\hat{a}\right)
\end{equation}
\begin{equation}\label{CSO3}
\hat{J_{z}}=\frac{1}{2}\left(\hat{b}^{\dagger}\hat{b} -\hat{a}^{\dagger}\hat{a}\right)
\end{equation}
in which the bosonic operators $\hat{a}^{\dagger}$ $(\hat{b}^{\dagger})$ and $\hat{a}$ $(\hat{b})$ denote the creation and annihilation operators for particles in $\left|\downarrow\right\rangle$ $(\left|\uparrow\right\rangle)$, respectively.

The three collective spin operators obey the commutation relation,
\begin{equation}\label{HUA}
[\hat{J}_{\alpha},\hat{J}_{\beta}] = i \hbar \epsilon_{\alpha\beta\gamma} \hat{J_{\gamma}}(\alpha,\beta,\gamma=x,y,z)
\end{equation}
where $\epsilon_{ijk}$ is the Levi-Civita symbol. Below, without loss of generality, we use the unit of $\hbar=1$. Thus, the collective spin operators obey the uncertainty relation,
\begin{equation}\label{HUE}
\Delta\hat{A}\Delta\hat{B} \geqslant \frac{1}{2}|\langle [\hat{A},\hat{B}] \rangle|,
\end{equation}
where $\Delta\hat{A}$ and $\Delta\hat{B}$ are standard deviations. Inserting Eq.~\eqref{HUA} into Eq.~\eqref{HUE}, one can obtain the uncertainty relation
\begin{equation}\label{HURS}
\Delta\hat{J}_{\alpha}\Delta\hat{J}_{\beta}\geqslant \frac{1}{2}|\langle \hat{J_{\gamma}}\rangle|,
\end{equation}
for the three collective spin operators.

Coherent spin states (CSS) are the most `classical like' pure quantum states of $N$ spin-$\frac{1}{2}$ particles that polarize in the same single-particle state~\cite{0022-3689-4-3-009}. Therefore, an arbitrary CSS can be expressed as
\begin{equation}\label{CSS1}
|\theta,\varphi\rangle_{CSS}=\otimes_{l=1}^{N} \left[sin(\theta/2)e^{-i\varphi/2}|\uparrow\rangle_{l} +cos(\theta/2)e^{i\varphi/2}|\downarrow\rangle_{l}\right],
\end{equation}
which can be generated from the all-spin-down state $\left|0,0\right\rangle_{CSS}=\otimes_{l=1}^{N} \left|\downarrow\right\rangle_{l}$ by a unitary rotation with angles $\theta$ and $\varphi$. The direction from the origin to the point $(\theta,\varphi)$ on the Bloch sphere corresponds to the direction of the mean total spin $\left\langle J\right\rangle$, which is called the mean spin direction (MSD).

One can also express a CSS in terms of the Dicke basis $|J,m\rangle$. The Dicke states are defined by the common eigenstates of $\hat{J}^2$ and $\hat{J}_{z}$: $\hat{J}^2|J,m\rangle=J(J+1)|J,m\rangle$ and $\hat{J}_{z}|J,m\rangle=m|J,m\rangle$. Here, we have $J=N/2$ and $-N/2\leqslant m\leqslant N/2$ with the total particle number $N$~\cite{PhysRevA.6.2211}. A general form of CSS in the Dicke basis reads as~\cite{RevModPhys.62.867,PhysRevA.68.033821}
\begin{equation}\label{CSS3}
|\theta,\varphi\rangle_{CSS} = \displaystyle \sum_{m=-J}^{J} C_{m}(\theta) e^{-i(J+m)\varphi} |J,m\rangle,
\end{equation}
with the coefficients
\begin{eqnarray}
C_{m}(\theta) &=& \left(
\begin{array}{cc}
2J  \\
J+m  \\
\end{array}
\right)^{\frac{1}{2}} \cos^{J-m}(\theta/2) \sin^{J+m}(\theta/2) \nonumber\\
&=& \left[\frac{(2J)!}{(J+m)!(J-m)!}\right]^{\frac{1}{2}} \cos^{J-m}(\theta/2) \sin^{J+m}(\theta/2),
\end{eqnarray}
which is a binomial distribution.

One important feature of CSS is that all particles are independent and have no quantum correlations. Therefore, a CSS has equal variance $(\Delta\hat{J}_{\bot})^{2}$ in any direction $J_{\bot}$ orthogonal to the MSD $(\theta,\varphi)$. The variance $(\Delta\hat{J}_{\bot})^{2}$ is given by the sum of $N$ variances $(\Delta\hat{S}_{\bot})^{2}$ of individual spin-$\frac{1}{2}$ particles, that is,
\begin{equation}
(\Delta\hat{J}_{\bot})^{2}=N\times(\Delta\hat{S}_{\bot})^{2} =\frac{N}{4}.
\end{equation}
Choosing the MSD along the $z$-axis, and the two orthogonal directions to the MSD along the $x$-axis and $y$-axis, we have
\begin{equation}
(\Delta\hat{J}_{x})^{2}=(\Delta\hat{J}_{y})^{2} =\frac{\left|\left\langle J_{z}\right\rangle \right|}{2}=\frac{N}{4},
\end{equation}
and
\begin{equation}
\Delta\hat{J}_{x}\Delta\hat{J}_{y}=\frac{\left|\left\langle J_{z}\right\rangle \right|}{2}.
\end{equation}
This indicates that the CSS satisfies the minimal condition of the Heisenberg uncertainty relation, i.e., Eq.~\eqref{HUA} takes the equal sign.

Similar to the state for a single spin-$\frac{1}{2}$ particle,  $|\theta,\varphi\rangle_{CSS}$ can also be represented on a generalized Bloch sphere with a radius of the total spin length $J = N/2$. Given the polar angle $\theta$ and the azimuthal angle $\varphi$ for a CSS, we have $\langle\hat{J}_{x}\rangle = J\cos\varphi\sin\theta$ and $\langle\hat{J}_{y}\rangle= J\sin\varphi\sin\theta$ and $\langle\hat{J}_{z}\rangle=J\cos\theta$.

The CSS can be used to perform the Ramsey interferometry~\cite{PhysRevA.82.045601}. The procedure of the Ramsey interferometry with a CSS is similar to the case of a single-particle state. Usually, the interferometry experiment starts from an initial state of all $N$ atoms in the same internal state $\left|\downarrow\right\rangle$, that is, $\left|\Psi_{0}\right\rangle = \otimes_{l=1}^N \left|\downarrow\right\rangle_{l}$. Applying a $\frac{\pi}{2}$ pulse to couple the two involved internal states, one can obtain the CSS,
\begin{equation}\label{Ramsey_CSS1}
|\Psi_{in}\rangle=\otimes_{l=1}^{N}\left[\frac{1}{\sqrt{2}} \left(\left|\uparrow\right\rangle_{l} +\left|\downarrow\right\rangle_{l}\right)\right]
\end{equation}
which has all atoms in the equal superposition of two internal states. Obviously, $\left|\Psi_{in}\right\rangle$ is a CSS with $\theta=\frac{\pi}{2}$ and $\varphi=0$. Then, the system undergoes a free evolution for a period of time, in which the ground state $\left|\downarrow\right\rangle$ accumulates a phase $\varphi/2$ and the excited state $\left|\uparrow\right\rangle$ accumulates a phase $-\varphi/2$. At the end of the free evolution, the state reads as,
\begin{equation}\label{Ramsey_CSS2}
\left|\Psi_{out}\right\rangle=\otimes_{l=1}^{N} \left[\frac{1}{\sqrt{2}} \left(e^{-i\frac{\varphi}{2}}\left|\uparrow\right\rangle_{l} +e^{+i\frac{\varphi}{2}} \left|\downarrow\right\rangle_{l}\right)\right].
\end{equation}
Finally, the second $\frac{\pi}{2}$ pulse is applied and the accumulated relative phase is extracted from the mean population difference $\left\langle n_{\uparrow}-n_{\downarrow}\right\rangle = 2 \left\langle \hat{J}_{z}\right\rangle$.

The evolution from the initial state $|\Psi_{0}\rangle$ to the final state $|\Psi_{f}\rangle$ is described by
\begin{equation}
|\Psi_{f}\rangle = \hat{U} |\Psi_{0}\rangle,
\end{equation}
with the propagation operator~\cite{PhysRevA.50.67,PhysRevA.57.4004}
\begin{equation}\label{PO_CSS}
\hat{U} = \exp\left(-i\frac{\pi}{2}\hat{J}_{y}\right) \exp\left(-i \varphi \hat{J}_{z}\right) \exp\left(-i\frac{\pi}{2}\hat{J}_{y}\right).
\end{equation}
Therefore, the expectation value for the final state is given by
\begin{eqnarray}\label{Jz_CSS}
\langle \hat{J}_{z} \rangle _{f}  =  \langle \Psi_{f}|\hat{J}_{z}|\Psi_{f}\rangle
=\langle\Psi_{0}|\hat{U}^{\dagger}\hat{J}_{z} \hat{U}|\Psi_{0}\rangle.
\end{eqnarray}
Substituting Eq.~\eqref{PO_CSS} into Eq.~\eqref{Jz_CSS}, we get
\begin{equation}\label{Jzf_CSS}
\langle \hat{J}_{z} \rangle_{f} = -\cos \varphi \langle \hat{J}_{z} \rangle_{0} + \sin \varphi \langle \hat{J}_{y} \rangle_{0},
\end{equation}
and
\begin{eqnarray}\label{DeltaJzf_CSS}
 \left(\Delta \hat{J}_{z}\right)^2_{f}
& = & \left(\Delta \hat{J}_{z}\right)^2_{0} \cos^2 \varphi +\left(\Delta \hat{J}_{y}\right)^2_{0} \sin^2 \varphi\nonumber\\
& & - \sin \varphi \cos \varphi \left\langle \hat{J}_{z}\hat{J}_{y}+\hat{J}_{y}\hat{J}_{z} \right\rangle.
\end{eqnarray}
As the initial state $|\Psi_{0}\rangle = \otimes_{l=1}^N |\downarrow\rangle_{l}$, we have
\begin{equation}
\langle \hat{J}_{z} \rangle_{0}=-\frac{N}{2}, \left(\Delta \hat{J}_{z}\right)_{0}=0,
\left(\Delta \hat{J}_{x}\right)_{0}=\left(\Delta \hat{J}_{y}\right)_{0}=\frac{\sqrt{N}}{2}.
\end{equation}
Thus the standard deviation of the final state reads as,
\begin{equation}
\left(\Delta \hat{J}_{z}\right)_{f} = \frac{\sqrt{N}}{2} \sin \varphi.
\end{equation}
Applying Eq.~\eqref{epf}, the standard deviation of $\varphi$ is given as
\begin{equation}
\Delta \varphi = \frac{( \Delta \hat{J}_{z} ) _{f}}{\left|\partial\langle \hat{J}_{z} \rangle _{f} /\partial\varphi\right|} = \frac{1}{\sqrt{N}}.
\end{equation}
This means that the measurement precision for the Ramsey interferometry with CSS obeys the scaling imposed by the standard quantum limit.

\begin{figure}[htb]
\centering
\includegraphics[width=4in]{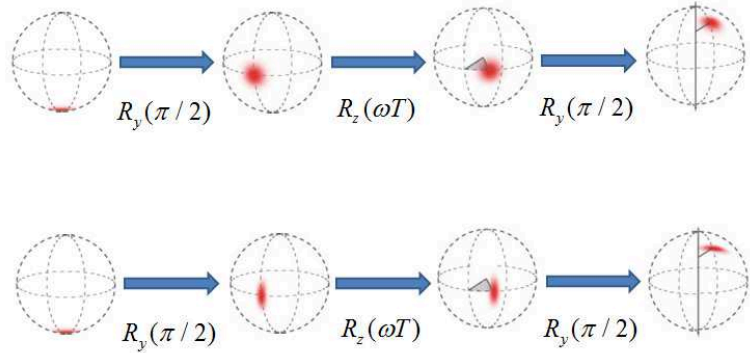}\caption{Schematic diagrams  of Ramsey interferometry on the Bloch sphere. Top: The initial state is a coherent spin state. Bottom: The initial state is a spin squeezed state. Adapted from Ref.~\citen{Gross2010}.}
\label{Fig3}
\end{figure}

\subsection{Spin squeezed states}

Similar to the quantum squeezing of position and momentum,  without violating the Heisenberg uncertainty relation, the fluctuations of one spin component can be reduced below the symmetric limit at the expense of the increased fluctuations of the other spin component. The state of reduced spin fluctuations along a specific direction is called spin squeezed state (SSS)~\cite{Walls1983,Sackett2010}. The reduced spin fluctuations may be employed to increase the measurement precision. The occurrence of quantum spin squeezing relates to quantum entanglement among the particles~\cite{PhysRevLett.96.050503,PhysRevA.68.012101,PhysRevLett.99.170405,PhysRevLett.86.4431}.

There are several definitions for quantum spin squeezing~\cite{Ma2011}. The first squeezing parameter $\xi_{H}$ is defined according to Heisenberg uncertainty relation~\cite{PhysRevLett.47.709}. It can be expressed as
\begin{equation}
\xi_{H}^{2}=\frac{2(\Delta\hat{J_{\alpha}})^{2}} {\left|\left\langle \hat{J_{\gamma}}\right\rangle \right|},\alpha\neq\gamma\in(x,y,z).
\end{equation}
A state of $\xi_{H}^{2}<1$ is a spin squeezed state. The second squeezing parameter $\xi_{S}$ is defined by the minimum fluctuation along the direction perpendicular to the MSD~\cite{PhysRevA.47.5138}. It can be written as
\begin{equation}
\xi_{S}^{2}=\frac{\textrm{min} (\Delta\hat{J}_{\vec{n}_{\bot}})^{2}}{j/2} =\frac{4\textrm{min}(\Delta\hat{J}_{\vec{n}_{\bot}})^{2}}{N}.
\end{equation}
Here, the minimization over all possible directions $\vec{n}_{\bot}$ is to find the most squeezed direction perpendicular to the MSD. The state is supposed to be squeezed if $\xi_{S}^{2}<1$. The third squeezing parameter $\xi_{R}$ is defined by the ratio of the phase fluctuation for the considered state and a reference CSS~\cite{PhysRevA.46.R6797,PhysRevA.50.67,PhysRevA.64.053815}. It reads as
\begin{equation}
\xi_{R}^{2}=\frac{\Delta\phi}{(\Delta\phi)_{css}} =\frac{N(\Delta\hat{J}_{\vec{n}_{\bot}})^{2}}{\left|\left\langle \hat{J}\right\rangle \right|^{2}}.
\end{equation}
This spin squeezing parameter is widely used in atomic Ramsey interferometry. In realistic experiments, $\Delta\hat{J}_{\vec{n}_{\bot}}$ can be obtained by measuring the population difference after an appropriate state rotation and $\left|\left\langle\hat{J}\right\rangle\right|$ can be extracted from the Ramsey fringes contrast. The state is spin squeezed if $\xi_{R}^{2}<1$. There are small differences among the definitions for the spin squeezing parameter. For an example, the differences between the definitions (83) and (84) are analyzed and the source of differences is explained by using the negativity criterion for entanglement~\cite{PhysRevA.68.064301}.

It has demonstrated that the Ramsey interferometry with a SSS as the input state may have a higher measurement precision than the case of a CSS~\cite{PhysRevA.64.052106}. Here, we assume the initial state is a SSS of
\begin{equation}\label{DeltaJy_SSS}
\left\langle \hat{J}_{y} \right\rangle_{0}=0 , \left\langle \hat{J}_{z}\right\rangle_{0} =-\frac{N}{2} , \left(\Delta \hat{J}_{y}\right)_{0} = \frac{\sqrt{N}}{2} \xi_{R},
\end{equation}
where the initial spin fluctuations are squeezed along the $y$-axis, i.e., the squeezing parameter $\xi_{R}<1$.

Similar to the single-particle case, the interferometry with a SSS also includes two $\frac{\pi}{2}$ pulses and a free evolution. The first $\frac{\pi}{2}$ pulse rotates the initial SSS an angle $\frac{\pi}{2}$ around the $y$-axis. In the free evolution process, the state rotates around the $z$-axis with an unknown angle $\varphi$ to be measured. Then applying the second $\frac{\pi}{2}$ pulse, the state rotates another $\frac{\pi}{2}$ around the $y$-axis. Finally, an additional rotation around its center is applied before measuring the population difference, which ensures the final population difference has minimal fluctuations~\cite{PhysRevA.46.R6797}.

The evolution from the initial state to the final state can be written as
\begin{equation}
|\psi_{f}\rangle=\hat{U}|\psi_{0}\rangle,
\end{equation}
with the propagation operator
\begin{equation}\label{PO_SSS}
\hat{U} = \exp(-i\frac{\pi}{2}\hat{J}_{y}) \exp(-i\varphi\hat{J}_{z})\exp(-i\frac{\pi}{2}\hat{J}_{y}).
\end{equation}
The expectation value of $\hat{J}_{z}$ is given as
\begin{equation}\label{Jzf_SSS}
\langle \hat{J}_{z} \rangle_{f} = -\cos \varphi \langle \hat{J}_{z} \rangle_{0} + \sin \varphi \langle \hat{J}_{y} \rangle_{0},
\end{equation}
and its variance reads as
\begin{eqnarray}\label{DeltaJzf_CSS}
 \left(\Delta \hat{J}_{z}\right)^2_{f}
&=& \left(\Delta \hat{J}_{z}\right)^2_{0} \cos^2 \varphi +\left(\Delta \hat{J}_{y}\right)^2_{0} \sin^2 \varphi\nonumber\\
& & - \sin \varphi \cos \varphi \left\langle \hat{J}_{z}\hat{J}_{y}+\hat{J}_{y}\hat{J}_{z} \right\rangle.
\end{eqnarray}
Obviously, the final state $|\psi_{f}\rangle$ is a SSS where the squeezed direction forms at an angle $\alpha$ relative to the $\hat{J}_{x}$ axis. To read out $\hat{J}_{z}$ with minimal uncertainty, one has to rotate the uncertainty ellipse of $|\psi_{f}\rangle$ around its center by an angle related to $\alpha$. Therefore, without any change of the expectation value of $\hat{J}_{z}$, the readout variance of $\hat{J}_{z}$ reads as
\begin{equation}
\left(\Delta \hat{J}_{z}\right)^2_{\textrm{readout}} = \left(\Delta \hat{J}_{y}\right)^2_{0}.
\end{equation}
which is given by the initial variance of $\hat{J}_{y}$. According to Eq.~\eqref{epf}, the standard deviation of $\varphi$ can be expressed as,
\begin{equation}
\Delta \varphi = \frac{( \Delta \hat{J}_{z} ) _{0}}{\left|\partial\langle \hat{J}_{z} \rangle _{f} /\partial\varphi\right|} =
\frac{\xi_{R}\sqrt{N}/2}{|\sin \varphi| N/2}=\frac{\xi_{R}}{|\sin \varphi| \sqrt{N}}.
\end{equation}
Clearly, the measurement precision $\Delta \varphi$ reaches its minimum,
\begin{equation}
\Delta \varphi =\frac{\xi_{R}}{\sqrt{N}}
\end{equation}
if $\varphi=\frac{\pi}{2}$. In comparison to coherent spin states, spin squeezed states may be used to beat the standard quantum limit: $\Delta \varphi \sim 1/\sqrt{N}$. For an example, if the initial state is squeezed to $\xi_{R} \sim 1/\sqrt{N}$, the best measurement precision can reach the Heisenberg limit: $\Delta \varphi \sim 1/N$.

\subsection{NOON states}

Theoretically, the NOON state has been proposed as one of the best candidates to improve the measurement precision. An NOON state is an equal-probability superposition of all $N$ particles in mode $a$ with zero particle in mode $b$, and vice versa. If the two modes are regarded as two possible paths for particles, the NOON state can be interpreted as all $N$ particles pass through either path $a$ or path $b$ together, which is also called the path-entangled state~\cite{PhysRevLett.85.2733, leea2002, PhysRevA.84.043628, Banaszek2009}. In general, it can be written in form of
\begin{equation}
|\textrm{NOON}\rangle=\frac{1}{\sqrt{2}} \left(|N\rangle_{a}|0\rangle_{b} +e^{i\theta}|0\rangle_{a}|N\rangle_{b}\right),
\end{equation}
with $\theta$ denoting an arbitrary phase. The NOON state is equivalent to the $N$-particle GHZ state, which is the maximally entangled state for a multiparticle system of two-state particles~\cite{Zhengfeng2008,PhysRevA.76.032111,PhysRevA.85.035601}. For a multiparticle system involving two single-particle states $\left|\downarrow\right\rangle$ and $\left|\uparrow\right\rangle$, the $N$-particle GHZ state can be expressed as
\begin{equation}
|\textrm{GHZ}\rangle=\frac{1}{\sqrt{2}} \left(|\frac{N}{2},+\frac{N}{2}\rangle +e^{i\theta}|\frac{N}{2},-\frac{N}{2}\rangle\right),
\end{equation}
with the Dicke basis, $\left|J=\frac{N}{2}, m=\frac{1}{2}(n_{\uparrow}-n_{\downarrow})\right\rangle$.

Below, we consider the Ramsey interferometry with,
\begin{equation}
|\Psi\rangle_{in} = \frac{1}{\sqrt{2}} \left(|\frac{N}{2},+\frac{N}{2}\rangle +|\frac{N}{2},-\frac{N}{2}\rangle\right),
\end{equation}
as the input state before the free evolution. In the free evolution, because of the entanglement, all particles simultaneously acquire the phase shift and each particle contributes a phase shift $+\frac{\varphi}{2}$ (or $-\frac{\varphi}{2}$) corresponding to $\left|\downarrow\right\rangle$ (or $\left|\uparrow\right\rangle$). Therefore, after the free evolution, the state reads as
\begin{equation}
|\Psi\rangle_{out} = \frac{1}{\sqrt{2}} \left(e^{-i\frac{N\varphi}{2}}|\frac{N}{2},+\frac{N}{2}\rangle +e^{+i\frac{N\varphi}{2}}|\frac{N}{2},-\frac{N}{2}\rangle\right).
\end{equation}

Below, we employ the quantum Fisher information to analyze the minimal uncertainty of measuring the phase $\varphi$. By differentiating the output state $|\Psi\rangle_{out}$ with respect to the relative phase $\varphi$, we have
\begin{equation}
\frac{d|\Psi\rangle_{out}}{d\varphi} =-\frac{iN}{2\sqrt{2}} \left(e^{-i\frac{N\varphi}{2}}|\frac{N}{2},+\frac{N}{2}\rangle - e^{+i\frac{N\varphi}{2}}|\frac{N}{2},-\frac{N}{2}\rangle\right).
\end{equation}
According to Eq.~\eqref{FI4}, the quantum Fisher information is given as
\begin{equation}
F_{Q}^{N}=4\left[\langle\psi'(\varphi)|\psi'(\varphi)\rangle -\left|\langle\psi'(\varphi)|\psi(\varphi)\rangle\right|^{2}\right] =4\left(\frac{N^2}{4}-0\right)=N^{2}.
\end{equation}
Therefore, the phase uncertainty satisfies,
\begin{equation}
\Delta\varphi\geq \frac{1}{F_{Q}}=\frac{1}{N},
\end{equation}
which is the scaling of the Heisenberg limit. In comparison to the case of independent particles, the measurement precision is improved from the standard quantum limit to the Heisenberg limit~\cite{PhysRevLett.71.1355}. However, in realistic experiments, it is hard to prepare a large-N GHZ state and the GHZ state is fragile in the presence of particle losses~\cite{Afek14052010}.

\subsection{Entangled coherent states}

An entangled coherent state (ECS) is a superposition of multimode coherent states~\cite{1751-8121-45-24-244002,PhysRevA.45.6811}. A typical class of the ECS is defined as~\cite{PhysRevLett.107.083601},
\begin{eqnarray}\label{ECS}
|\textrm{ECS}\rangle &=& e^{(-|\alpha|^2)/2} N_{\alpha} \displaystyle \sum_{n=0}^{\infty} \frac{\alpha^n}{n!}\left[(\hat{a}^{\dagger})^n + (\hat{b}^{\dagger})^n\right] |0\rangle_{a} |0\rangle_{b} \nonumber\\
&=& N_{\alpha} \left[|\alpha \rangle_{a}|0 \rangle_{b} + |0 \rangle_{a}|\alpha \rangle_{b} \right],
\end{eqnarray}
with the normalization factor $N_{\alpha}=1/\sqrt{2(1 + e^{-|\alpha|^2})}$. This ECS can be understood as the superposition of multiple NOON states with different total particle numbers. As a coherent state involves Fock states with particle number from zero to infinity, the averaged total particle number but not the total particle number itself is a good quantity. The averaged total particle number of the ECS~\eqref{ECS} is given by
\begin{equation}
\langle n \rangle = \langle n_a + n_b \rangle= 2 N_{\alpha}^2 |\alpha|^2.
\end{equation}

Let us consider a Mach-Zehnder interferometry with the ECS~\eqref{ECS} as the input state for free propagation and the two modes $a$ and $b$ as two paths. Here, we assume that each particle in mode $b$ acquires a phase shift $\varphi$ with respect to the one in mode $a$. That is, the state after the free propagation reads as
\begin{eqnarray}
|\Psi\rangle_{out}
& = & N_{\alpha} e^{-\frac{|\alpha|^2}{2}} \displaystyle \sum_{n=0}^{\infty} \frac{\alpha^n}{n!}\left[(\hat{a}^{\dagger})^n + (\hat{b}^{\dagger})^n e^{i n\varphi}\right] |0\rangle_{a} |0\rangle_{b} \\
& = & N_{\alpha} e^{-\frac{|\alpha|^2}{2}} \left[ \left( \displaystyle \sum_{n=0}^{\infty} \frac{\alpha^n}{\sqrt{n!}}|n\rangle_{a}\right) |0\rangle_{b} + \left(\displaystyle \sum_{n=0}^{\infty} \frac{\alpha^n}{\sqrt{n!}}e^{i n \varphi}|n\rangle_{b}\right)|0\rangle_{a} \right]\nonumber
\end{eqnarray}
Therefore, the derivative of the output state $|\Psi\rangle_{out}$ with respect to $\varphi$ reads as
\begin{eqnarray}
\frac{d|\Psi\rangle_{out}}{d\varphi}
& = & N_{\alpha} \left[ |0\rangle_{a} \left( e^{-\frac{|\alpha|^2}{2}}\displaystyle \sum_{n=0}^{\infty} \frac{ i n \alpha^n}{\sqrt{n!}}e^{i n \varphi}|n\rangle_{b}\right) \right].
\end{eqnarray}
According to Eq.~\eqref{FI4}, the quantum Fisher information is given by
\begin{equation}
F_{Q}=4\left|\alpha\right|^2 N_{\alpha}^2+ 4(1-N_{\alpha}^2)\left|\alpha\right|^4 N_{\alpha}^2.
\end{equation}
Thus the phase uncertainty $\Delta \varphi$ satisfies
\begin{equation}
\Delta \varphi \ge \frac{1}{2\left|\alpha\right| N_{\alpha} \sqrt{1+(1-N_{\alpha}^2)\left|\alpha\right|^2}}.
\end{equation}
If the parameters satisfy the conditions of $\alpha \gg 1$, $N_{\alpha} \approx 1/\sqrt{2}$ and $|\alpha|^2 \approx \langle n \rangle \gg 1$, the phase uncertainty obeys
\begin{equation}
\Delta \varphi \ge \frac{1}{\sqrt{2 \langle n \rangle} \sqrt{1+ \langle n \rangle/2}} \approx \frac{1}{\langle n \rangle}.
\end{equation}
This means that the phase uncertainty can approach to the Heisenberg limit.

\subsection{Twin Fock states}

The twin Fock state,
\begin{equation}\label{TWIN}
|\textrm{TWIN}\rangle=|N\rangle_{a} |N\rangle_{b},
\end{equation}
is a two-mode Fock state with equal particle number for the two modes. By using the twin Fock states as input states and parity measurements, it has been demonstrated the Heisenberg-limited Mach-Zehnder interferometry~\cite{PhysRevA.82.013831, PhysRevLett.89.150401, PhysRevA.68.023810, PhysRevA.70.033601}.

The first beam splitter transforms the twin Fock state into
\begin{equation}
|\Psi\rangle_{BS1}=\hat{U}_{BS1}|\textrm{TWIN}\rangle=\displaystyle \sum_{k=0}^N C_{k}^N |2k\rangle_{a}|2N-2k\rangle_{b},
\end{equation}
with the beam splitter operator,
\begin{equation}
\hat{U}_{BS1}=\exp\left[\frac{\pi}{4}(\hat{a}^{\dagger}\hat{b} -\hat{b}^{\dagger}\hat{a})\right],
\end{equation}
and the coefficients,
\begin{equation}
C_{k}^N=\frac{1}{2^N}(-1)^{N-k}\left[ \left(
            \begin{array}{cc}
              2k & \\
              k & \\
            \end{array}
          \right)   \left(
            \begin{array}{cc}
              2N-2k & \\
              N-k & \\
            \end{array}
          \right)   \right]^{1/2}.
\end{equation}
Here $\hat{a}$ and $\hat{b}$ are the two annihilation operators for particles in modes $a$ and $b$, respectively. A free propagation follows the first beam splitter, in which each particles in mode $b$ accumulates a relative phase $\varphi$. Therefore, after the free propagation, the state reads as
\begin{equation}
|\Psi(\varphi)\rangle=\displaystyle \sum_{k=0}^N e^{i\varphi(2N-2k)} C_{k}^N |2k\rangle_{a}|2N-2k\rangle_{b},
\end{equation}
which includes the information of the phase $\varphi$ to be measured. Then the second beam splitter,
\begin{equation}
\hat{U}_{BS2}=\exp\left[-i\frac{\pi}{4}(\hat{a}^{\dagger}\hat{b} +\hat{b}^{\dagger}\hat{a})\right],
\end{equation}
is applied to $|\Psi(\varphi)\rangle$ and the state becomes
\begin{equation}
|\Psi\rangle_{out}=\hat{U}_{BS2}|\Psi(\varphi)\rangle.
\end{equation}
At last, a parity measurement is performed for one of two modes. Parity measurements have been widely used to extract phase shifts in quantum optical metrology~\cite{Gerry2010} with highly entangled states including NOON states, entangled coherent states and twin Fock states. Further, the parity measurement has been adapted to extract the relative phase between Bose condensed atoms in different hyperfine levels~\cite{PhysRevA.68.025602}.

Here, the parity operator of mode $b$ can be expressed as
\begin{equation}
\hat{\Pi}_{b}=\exp(i\pi\hat{b}^{\dagger}\hat{b}),
\end{equation}
and its expectation value is a Legendre polynomial,
\begin{equation}
\langle \hat{\Pi}_{b} \rangle = \left\langle \Psi| \hat{\Pi}_{b}|\Psi\right\rangle_{out}= \left\langle \Psi(\varphi)|\hat{U}_{BS2}^{\dagger} \hat{\Pi}_{b} \hat{U}_{BS2}|\Psi(\varphi)\right\rangle = P_{N}[\cos(2\varphi)].
\end{equation}
According to Eq.~\eqref{epf}, the phase uncertainty is given by
\begin{equation}
\Delta \varphi = \frac{\Delta \hat{\Pi}_{b}} {\left|{\partial\langle\hat{\Pi}_{b}\rangle} /{\partial\varphi}\right|}.
\end{equation}
For $\varphi \rightarrow 0$, the phase uncertainty $\Delta \varphi$ versus the particle number $N$ approaches to the Heisenberg limit $\Delta \varphi_{HL}=1/(2N)$. For other values of $\varphi$, the phase uncertainty $\Delta \varphi$ will blow up for some specific values of the total particle number and it still approaches to the Heisenberg limit for the other values of the total particle number.

The phase uncertainty can also be derived by calculating the quantum Fisher information. The derivative of $|\Psi(\varphi)\rangle$ with respect to $\varphi$ reads as
\begin{equation}
\frac{d|\Psi(\varphi)\rangle}{d\varphi} =\displaystyle \sum_{k=0}^N i (2N-2k) e^{i\varphi(2N-2k)} C_{k}^N |2k\rangle_{a}|2N-2k\rangle_{b}.
\end{equation}
Substituting this derivative into Eq.~\eqref{FI4}, we obtain the quantum Fisher information,
\begin{equation}
F_{Q}=2N(1+N).
\end{equation}
Thus the quantum Cram\'{e}r-Rao bound for the phase uncertainty is given as
\begin{equation}
\Delta \varphi \ge \frac{1}{\sqrt{2N^2+2N}},
\end{equation}
which indicates that the minimal phase uncertainty obtained by the optimal measurement can reach the Heisenberg limit. This bound is consistent with the phase uncertainty obtained by the parity measurement.

\section{Experimental Progresses}

In recent years, there appear great advances in quantum metrology with multiparticle systems. In particular, multiparticle entangled states have been widely used to implement high-precision metrology from spectroscopy, interferometers to atomic clocks~\cite{LiYun2009,Bucker2011,PhysRevLett.108.130402,PhysRevA.86.063605,PhysRevA.84.023619}.To implement high-precision metrology with entangled multiparticle systems, in addition to the similar operations in quantum metrology with independent particles, a key problem is how to generate multiparticle entanglement. Usually, the multiparticle entanglement can be generated by intrinsic or artificial inter-particle interactions, such as, the intrinsic s-wave scattering between ultracold atoms, the Coulomb interaction between ultracold trapped ions, the laser-induced interaction and the continuous quantum non-demolition measurement. Below, we briefly review some typical progresses in quantum metrology with Bose-Einstein condensed atoms~\cite{Gross2010,Riedel2010,L¨¹cke11112011}, ultracold trapped ions~\cite{PhysRevLett.86.5870 ,Leibfried04062004} and cold atomic ensembles~\cite{12460641,Anne2010}.

\subsection{Bose-Einstein condensed atoms}

An atomic Bose-Einstein condensate (BEC) has intrinsic atom-atom interaction dominated by the s-wave scattering, which can be used to generate entangled states such as spin squeezed states~\cite{Julian2012,Orzel23032001,Esteve2008,
PhysRevA.81.063605,PhysRevA.71.042317,PhysRevA.66.033611}, GHZ states~\cite{PhysRevLett.90.030402}and twin Fock states~\cite{PhysRevA.70.033601,PhysRevA.82.013831}. In recent experiments, the spin squeezed states have been generated by one-axis twisting and then they are used to implement high-precision interferometry beyond the standard quantum limit~\cite{Gross2010,Riedel2010,PhysRevA.47.5138, PhysRevLett.101.040403,Jin2009,PhysRevLett.107.013601}. It has also been demonstrated the generation of twin Fock states via spin dynamics and the applications of the generated twin Fock states in high-precision interferometers~\cite{L¨¹cke11112011}. Different from the schemes with distinguishable particles, which need entangled input states for beating the standard quantum limit, the schemes with identical Bose condensed atoms do not need entangled input states~\cite{AnnPhys.325.924, JPhysB.44.091001, IJQI.9.1745} and the entanglement can be dynamically generated in these schemes via time evolution~\cite{Gross2010,Riedel2010}.

\subsubsection{Nonlinear interferometry with spin squeezed states}

Generally speaking, due to the intrinsic nonlinear interaction between atoms, almost all interferometers with Bose condensed atoms are nonlinear. Most of the interferometers with atomic BECs can be described by Bose-Josephson Hamiltonians. Here, for simplicity, we only discuss Bose-Josephson systems~\cite{Gati2007, Lee2012, Dalton2012, PhysRevLett.79.4950, PhysRevA.55.4318, PhysRevA.57.2920, PhysRevA.59.3868, PhysRevA.64.053604, PhysRevA.69.033611, PhysRevLett.97.150402, Lee2008, PhysRevlett.102.070401, PhysRevLett.106.120405, PhysRevA.80.053613, PhysRevA.80.013619, PhysRevA.84.023629, PhysRevA.80.023609, JPhysB.34.4689, PhysRevA.81.023615} of Bose-Einstein condensed atoms in two different modes. There are two typical Bose-Josephson systems. One is the external Bose-Josephson junction (BJJ), which is realized by an atomic BEC in a deep double-well potential. The other is the internal BJJ, which is realized by Bose-Einstein condensed atoms involving two coupled hyperfine states. There are lots of studies on macroscopic quantum phenomena in BJJs. Here, we concentrate our discussions on many-body quantum interferometry with quantized BJJs.

In second quantization, the external BJJ can be described by the many-body Hamiltonian,
\begin{eqnarray}\label{MBH1}
H &=& \int d\mathbf{r} \hat{\Psi}^{\dagger}(\mathbf{r}) \left[-\frac{\hbar^{2}\nabla^{2}}{2m} +V_{dw}(\mathbf{r})\right]\hat{\Psi}(\mathbf{r})\nonumber\\
 && +\frac{g}{2}\intop d\mathbf{r}\hat{\Psi}^{\dagger}(\mathbf{r})\hat{\Psi}^{\dagger} (\mathbf{r})\Psi(\mathbf{r})\hat{\Psi}(\mathbf{r}),
\end{eqnarray}
where $\hat{\Psi}(r)$ and $\hat{\Psi}^{\dagger}(r)$ are bosonic field operators, $V_{dw}(\mathbf{r})$ is the double-well potential, $g=4\pi\hbar a_{s}/m$, and $a_{s}$ is the s-wave scattering length. By applying the two-mode approximation~\cite{Dalton2012}, the field operator reads as,
\begin{equation}
\hat{\Psi}(\mathbf{r})=\hat{b}_{1}\phi_{1}(\mathbf{r}) +\hat{b}_{2}\phi_{2}(\mathbf{r}),
\end{equation}
with $\hat{b}_{1}$ and $\hat{b}_{2}$ represent the Bose annihilation operators for the atoms in the Wannier states $\phi_{1}(\mathbf{r})$ and $\phi_{2}(\mathbf{r})$, respectively.

While for an internal BJJ, it obeys the many-body quantum Hamiltonian,
\begin{equation}
H=\int d\mathbf{r}
\left(\hat{\Psi}^{\dagger}_{1}(\mathbf{r}), \hat{\Psi}^{\dagger}_{2}(\mathbf{r})\right)
\left(
\begin{array}{cc}
h_{1}^{(0)} & ~~-\frac{\hbar\Omega}{2} \\
-\frac{\hbar\Omega}{2} & ~~h_{2}^{(0)} \\
\end{array}
\right)
\left(
\begin{array}{cc}
\hat{\Psi}_{1}(\mathbf{r}) & \\
\hat{\Psi}_{2}(\mathbf{r}) & \\
\end{array}
\right) + H_{int},
\end{equation}
with
\begin{eqnarray}
h_{1}^{(0)} &=& -\frac{\hbar^{2}\nabla^{2}}{2m} +V_{1}(\mathbf{r})-\frac{\Delta}{2},\nonumber\\
h_{2}^{(0)} &=& -\frac{\hbar^{2}\nabla^{2}}{2m} +V_{2}(\mathbf{r})+\frac{\Delta}{2},\nonumber
\end{eqnarray}
and
\begin{eqnarray}
H_{int} &=& H_{11}+H_{22}+H_{12},\nonumber\\
H_{11} &=& \frac{g_{11}}{2}\intop d\mathbf{r}\hat{\Psi}_{1}^{\dagger}(\mathbf{r}) \hat{\Psi}_{1}^{\dagger} (\mathbf{r}) \Psi_{1}(\mathbf{r})\hat{\Psi}_{1}(\mathbf{r}),\nonumber\\
H_{22} &=& \frac{g_{22}}{2}\intop d\mathbf{r}\hat{\Psi}_{2}^{\dagger}(\mathbf{r}) \hat{\Psi}_{2}^{\dagger} (\mathbf{r}) \Psi_{2}(\mathbf{r})\hat{\Psi}_{2}(\mathbf{r}),\nonumber\\
H_{12} &=& \frac{g_{12}}{2}\intop d\mathbf{r}\hat{\Psi}_{2}^{\dagger}(\mathbf{r}) \hat{\Psi}_{1}^{\dagger} (\mathbf{r}) \Psi_{1}(\mathbf{r})\hat{\Psi}_{2}(\mathbf{r}).\nonumber
\end{eqnarray}
Here, $\Omega$ is the Rabi frequency of the coupling, $\Delta$ is the detuning to resonance, $V_{j}(\mathbf{r})$ denote the trapping potentials, and $g_{ij}$ describe the s-wave scattering of atoms in modes $i$ and $j$. For a spin-independent trap $V_{1}(\mathbf{r})=V_{2}(\mathbf{r})$, assuming all atoms staying in the same spatial state $\phi(\mathbf{r})$, we can apply the two-mode approximation,
\begin{equation}
\hat{\Psi}_{j}(\mathbf{r})=\hat{b}_{j}\phi(\mathbf{r}), ~~(j=1 \textrm{~and~} 2),
\end{equation}
with $\hat{b}_{1}$ and $\hat{b}_{2}$ being the annihilation operators for the atoms in the two hyperfine states.

By integrating the spatial coordinates, both external and internal BJJs can be described by a unified two-mode Bose-Hubbard model~\cite{Lee2012},
\begin{equation}\label{BHH}
H=-\frac{J}{2}\left(\hat{b}_{2}^{\dagger}\hat{b}_{1} +\hat{b}_{1}^{\dagger}\hat{b}_{2}\right) +\frac{\delta}{2}\left(\hat{n}_{2}-\hat{n}_{1}\right) +\frac{E_{c}}{8}\left(\hat{n}_{2}-\hat{n}_{1}\right)^{2},
\end{equation}
where $J$ is the tunneling strength, $\delta$ is the imbalance and $E_{c}$ is the effective ``charging" energy. For an external BJJ, $E_{c} \propto g$. While for an internal BJJ, $E_{c} \propto g_{11}+g_{22}-2g_{12}$. Obviously, $\left[\hat{N},H\right]=0$, therefore the total atomic number $\hat{N}=\hat{b}_{1}^{\dagger}\hat{b}_{1} +\hat{b}_{2}^{\dagger}\hat{b}_{2}$ is conserved. By using the Fock basis $\left\{|n_{1},n_{2}\rangle\right\}$, an arbitrary state can be expressed as $|\Psi\rangle=\Sigma_{n_{1},n_{2}} C_{n_{1}n_{2}}|n_{1},n_{2}\rangle$,
where $n_{j}=\hat{b}_{j}^{\dagger}\hat{b}_{j}$ are the number of particles in the $j$-th mode.

Ground state and quasi-particle excitations of the BJJ Hamiltonian~\eqref{BHH} sensitively depend on the ratio between the tunneling strength $J$ and the charging energy $E_c$~\cite{JPhysB.34.4689, Lee2008, Gati2007, PhysRevA.81.023615}. The competition between the Josephson tunneling and the nonlinear interaction results in different ground state behaviors. For a symmetric BJJ ($\delta=0$), dependent on the ratio $\left|E_{c}/J\right|$, the system shows three different regimes: Rabi, Fock and Josephson regimes~\cite{Gati2007}.
\begin{itemize}
\item \textbf{(a) Rabi regime}, $\left|E_{c}/J\right|\ll N^{-1}$, in which the system is dominated by the Josephson term of $J$ and there is a well defined relative phase between the two modes. The ground state is a coherent spin state, $|CSS\rangle=\exp(i\varphi \hat{J}_{z})\exp(i\theta \hat{J}_{y})|N/2,+N/2\rangle$.
\item \textbf{(b) Fock regime}, $\left|E_{c}/J\right|\gg N$, in which the system is dominated by the nonlinear interaction term of $E_c$ and the relative phase between the two modes is completely random. The ground state depends on the sign of the nonlinear interaction term. If $E_{c}$ is positive, the ground state is a single Fock state $|\frac{N}{2},\frac{N}{2}\rangle$ for even $N$ or a superposition of two Fock states $\frac{1}{\sqrt{2}}\left(|\frac{N}{2},-1\rangle +|\frac{N}{2},1\rangle\right)$ for odd $N$. If $E_{c}$ is negative, there are two degenerate ground states $|\frac{N}{2},\frac{N}{2}\rangle$ and $|\frac{N}{2},-\frac{N}{2}\rangle$. Therefore, any superposition of these two degenerated states including the GHZ state $\frac{1}{\sqrt{2}}\left(|\frac{N}{2},\frac{N}{2}\rangle +|\frac{N}{2},-\frac{N}{2}\rangle\right)$ is also a ground state. It has been proposed that the GHZ state can be adiabatically prepared and then it can be used to achieve a Heisenberg-limited interferometry~\cite{PhysRevLett.97.150402}.
\item \textbf{(c) Josephson regime}, $N^{-1}\ll\left|E_{c}/J\right|\ll N$, in which the number imbalance and the relative phase are both fluctuating and the ground state is an intermediate squeezed state.
\end{itemize}
In terms of the collective spin operators \eqref{CSO1}, \eqref{CSO2} and \eqref{CSO3}, the BJJ Hamiltonian~\eqref{BHH} becomes as
\begin{equation}
H=-B_{x}\hat{J}_{x}+B_{z}\hat{J}_{z}+\chi\hat{J}_{z}^{2},
\end{equation}
where the transverse magnetic field $B_{x}=2J$, the longitudinal magnetic field $B_{z}=\delta$ and the nonlinear interaction energy $\chi=\frac{E_{c}}{2}$. This Hamiltonian is just the `one-axis twisting' Hamiltonian for generating quantum spin squeezing~\cite{PhysRevA.47.5138}. In recent years, several theoretical methods for generating spin squeezed states in BJJs have been proposed~\cite{PhysRevA.82.063622, PhysRevA.84.022107, PhysRevA.86.023615, PhysRevA.86.023623}. In realistic systems of Bose condensed atoms, thermal atoms and atom loss have significant effects on the achievable optimal spin squeezing~\cite{PhysRevA.82.045601, PhysRevA.82.043624, PhysRevLett.100.210401, PhysRevLett.107.060404}.

Dramatic observations of phase fluctuations and number squeezing in external BJJs have been made. The phase fluctuations were observed by Gati et el.~\cite{PhysRevLett.96.130404}, Jo et al.~\cite{PhysRevLett.99.240406} and Hofferberth et al.~\cite{NatPhys.2.710, NatPhys.4.489}. The number squeezing was observed by Jo et al.~\cite{PhysRevLett.98.030407} and Esteve et al.~\cite{Nature.455.1216}. In addition, for an array of independent BECs, the phase fluctuations were observed by Hadzibabic et al.~\cite{PhysRevLett.93.180403} and the number squeezing was observed by Orzel et al.~\cite{Science.291.2386}. Moreover, the achieved squeezing in external BJJs, a kind of multiparticle entanglement, can be used to implement precision metrology with spatial atom interferometers~\cite{Nature.455.1216}.

The spin squeezing in internal BJJs and the use of the prepared spin squeezed states in precision metrology have also been demonstrated. Generally, the inter-mode coupling for an internal BJJ should be characterized by a Rabi frequency $\Omega$, a phase $\gamma$ and a detuning $\delta$, therefore the internal BJJ obeys the one-axis twisting Hamiltonian
\begin{equation}\label{OAT}
H=\Omega\hat{J}_{\gamma}+\delta\hat{J}_{z}+\chi\hat{J}_{z}^{2},
\end{equation}
with $\hat{J}_{\gamma}=(\cos\gamma)\hat{J}_{x} -(\sin\gamma)\hat{J}_{y}$. Described by this Hamiltonian, the spin squeezing of Bose condensed atoms has been demonstrated by two experimental groups: Oberthaler's group and Treutlein's group. They independently developed two different methods for turning on the strongly nonlinear interactions and then generating quantum spin squeezing. Oberthaler's group has used the Feshbach resonance to decrease the inter-component s-wave scattering length~\cite{Gross2010}. Taking the technical noises into account, the squeezing factor in this experiment~\cite{Gross2010} can reach $\xi_{N}^{2}=-8.2{}_{-1.2}^{+0.8}$ dB, which is close to the atom-loss-limited theoretical optimum for this system. Treutlein's group has used a state-dependent trap to decrease the density overlap between two components~\cite{Riedel2010}. To obtain the best squeezing angle in their experiments, they measured the squeezing factor for different rotation angle.

\subsubsection*{(a) The experiment with optical lattices}

By loading Bose condensed atoms into optical lattices, Oberthaler's group has successfully prepared the entanglement of about 170 $^{87}$Rb atoms and then realized a nonlinear Ramsey interferometer~\cite{Gross2010}. In comparison to the ideal phase sensitivity obtained by unentangled states, their experimental data show that the phase sensitivity is enhanced by 15 percent. They firstly prepare a BEC of $^{87}$Rb atoms occupying the hyperfine state $|F=1,m_{F}=-1\rangle$ in an optical dipole trap. Then, through supposing a one-dimensional optical lattice potential, the dipole trap splits into six, which allows to perform six independent experiments in parallel. Before applying the first $\frac{\pi}{2}$ pulse, the atoms are swept from the state $|F=1,m_{F}=-1\rangle$ to the state $|a\rangle=|F=1,m_{F}=1\rangle$. Since the first $\frac{\pi}{2}$ pulse, only two hyperfine states $|a\rangle=|F=1,m_{F}=1\rangle$ and $|b\rangle=|F=2,m_{F}=-1\rangle$ are involved and individual systems localized in each lattice site can be described by the one-axis twisting Hamiltonian. The effective nonlinear interaction $\chi\varpropto a_{aa}+a_{bb}-2a_{ab}$ relates to the intra-species and inter-species interactions. The inter-species interaction is tuned by the Feshbach resonance and $\chi=2\pi\times0.063$~Hz at a magnetic field of $B=9.10$~G. The Rabi frequency $\Omega$ can be switched rapidly from $0$ to $2\pi\times600$~Hz. Therefore, the system can be adjusted between Rabi regime and Fock regime.

\begin{figure}[htb]
\centering
\includegraphics[width=4in]{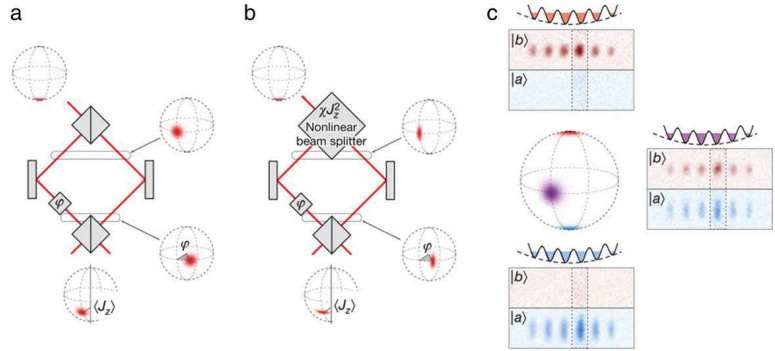}\caption{(a) Schematic of a classical linear interferometer. (b) Schematic of an nonlinear interferometer. (c) Six independent BECs of $^{87}$Rb in a one-dimensional optical lattice. Two-state atoms in each well form a two-mode system described by the one-axis twisting Hamiltonian. The individual detection of the condensate in each well can be achieved by high-intensity absorption imaging. From Ref.~\citen{Gross2010}.}
\label{Fig5}
\end{figure}

The input state for the interferometer is a coherent spin state polarized to the $\hat{J}_{z}$-direction. After the first $\frac{\pi}{2}$ pulse, the state rotates to the $\hat{J}_{x}$-direction, with $\left\langle J_{z}\right\rangle =\left\langle J_{y}\right\rangle =0$ and $\Delta J_{z}=\Delta J_{y}=\sqrt{N}/2$. Then the Josephson coupling $(\Omega J_{\gamma})$ is switched off, the system stays in the Fock regime and its state evolves under the nonlinear term, which induces a squeezing angle $\alpha_{0}$ with respect to z-direction. A rotation of the uncertainty ellipse around its center by $\alpha=\alpha_{0}+\pi/2$ is followed. Then the modes $|a\rangle$ and $|b\rangle$ experience a $\tau=2\mu s$ phase accumulation period and recombine via another $\frac{\pi}{2}$ pulse before the readout of population imbalance.

\subsubsection*{(b) The experiment with an atomic chip}

By loading Bose condensed atoms into an atomic chip, Treutlein's group has created spin-squeezed states which may improve the measurement precision beyond the standard quantum limit~\cite{Riedel2010}. In the experiment, two spin states $\left|F=1,m_{F}=-1\right\rangle$ and $\left|F=2,m_{F}=1\right\rangle$ of $^{87}$Rb atoms are involved and the system obeys the one-axis twisting Hamiltonian~\eqref{OAT}. The effective inter-component interaction is controlled by adjusting the spatial overlap between two spin components.

\begin{figure}[htb]
\centering
\includegraphics[width=4in]{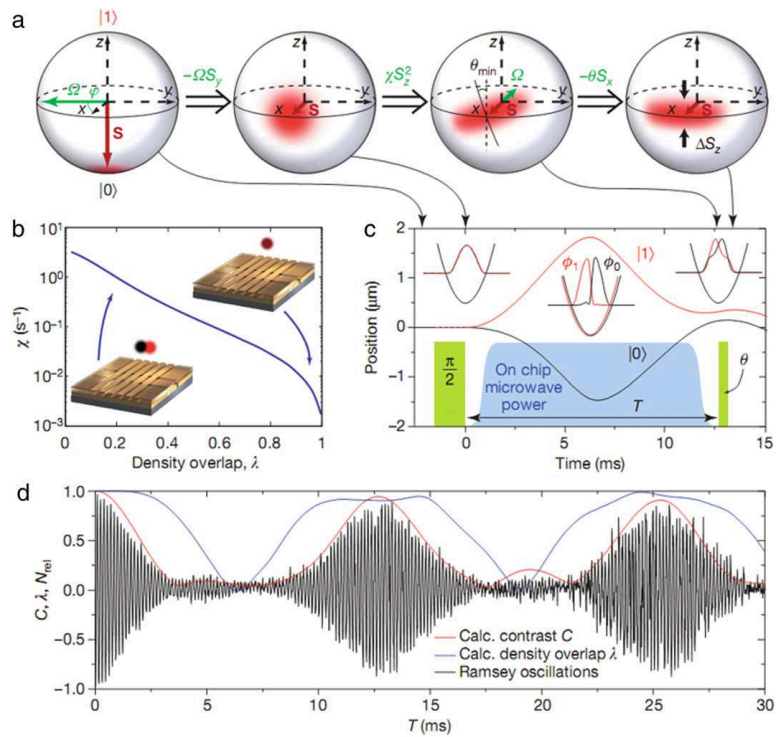}\caption{(a) The preparation of spin squeezing on the generalized Bloch sphere. (b) The nonlinearity $\chi$ is decreased as the increase of the normalized density overlap $\lambda$ of the two spin components. (c) The experimental sequence and the motion of two spin components corresponding to (a). (d) Measured Ramsey fringes in the normalized population difference $N_{rel}$. From Ref.~\citen{Riedel2010}.}
\label{Fig7}
\end{figure}

To prepare spin squeezing, except for controlling the nonlinear interaction, Treutlein's group has used similar procedures in the experiment of Oberthaler's group. First, a coherent spin state is prepared by a resonant $\frac{\pi}{2}$ pulse for 120 $\mu s$. During the pulse, the coupling term dominates, $\Omega\gg\chi N$, so that the atom-atom interaction can be neglected. The state-dependent microwave potential is turned on within 50 $\mu s$ to cause a sudden separation of trap minima for the two hyperfine states. The two components begin to oscillate oppositely, the overlap of the modes wavefunction reduces, which leads to the decreasing of the inter-component interaction and the increasing of effective nonlinearity $\chi$. The nonlinearity can attain $\chi=1.5$ $s^{-1}$ at the maximum separation. The two components overlap again after 12.7 $ms$ and the nonlinear interaction squeezing dynamics stops.

\subsubsection{Twin matter-wave interferometry}

In addition to the quantum interferometry with spin-squeezed states, twin matter-wave interferometry with initial twin Fock sates has been demonstrated in experiment~\cite{L¨¹cke11112011}. Different from the spin squeezing via one-axis twisting, spin exchange dynamics of Bose condensed atoms~\cite{PhysRevA.75.023605} has been used to create large ensembles of up to $10^{4}$ pair-correlated atoms from initial twin Fock states. Attribute to the pair correlation induced by spin exchange, the phase uncertainty can beaten the standard quantum limit.

The experiment starts with creating a $^{87}$Rb condensate of $2.8\times10^4$ atoms in the hyperfine state $|F=2,m_F=0\rangle$ in an optical dipole trap. Then the spin-exchange collision gradually produces correlated pairs of atoms with spins up and down. The functionality of the spin-exchange collision is just like a parametric amplifier, where the total number of the correlated pairs of atoms in $|F=2,m_F=\pm1\rangle$ increase exponentially with time. Afterwards, the trap is switched off and the three hyperfine states are split by a strong magnetic field gradient, and all three hyperfine states are recorded by absorption imaging. As the hyperfine states $|F=2,m_F=\pm1\rangle$ are generated in pairs, the number of particles in these two modes is exactly equal. Therefore, the twin Fock state is created with zero number difference and relative phase completely undetermined between these two modes.

\begin{figure}[htb]
\centering
\includegraphics[width=4in]{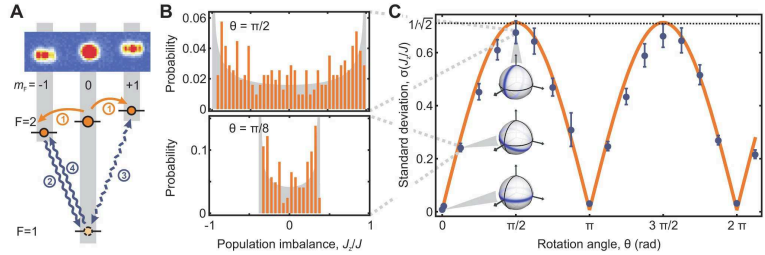}\caption{(A) Schematic of the sequence of the realization of the beam splitter. Three microwave pulses are sequentially applied to achieve the coupling of the two hyperfine states $|F=2,m_F=\pm1 \rangle$. The total effect is equivalent to a rotation around the $x$-axis by an angle $\theta$. (B) Distribution of the normalized population difference for two different $\theta$. The strongest fluctuations are obtained for $\theta=\frac{\pi}{2}$. The shaded area is the theoretical result. (C) The fluctuation of the normalized population difference versus the rotation angle $\theta$, where the standard deviation $\sigma(J_z/J)$ oscillates approximately as $\sigma(J_z/J)=|\sin \theta|/\sqrt{2}$. From Ref.~\citen{L¨¹cke11112011}.}
\label{Fig10}
\end{figure}

Then the generated twin Fock state is input for implementing interferometry. The beam splitter of the interferometer is realized by three resonant microwave pulses. The first one is applied to transfer the atoms in $|F=2,m_F=-1\rangle$ to $|F=1,m_F=0\rangle$. The second pulse with duration $\tau$ couples the states $|F=1,m_F=0\rangle$ and $|F=2,m_F=1\rangle$. The third pulse transfers the atoms from $|F=1,m_F=0\rangle$ to $|F=2,m_F=-1\rangle$. The action of these three pulses is equivalent to a rotation around the $x$-axis by an angle $\theta=\tau \Omega_R$, where $\Omega_R$ is the Rabi frequency and $\tau$ is the duration of the coupling pulse. For $\theta=\frac{\pi}{2}$, the fluctuation of the population imbalance is maximal and it corresponds to the $\frac{\pi}{2}$ pulse in a Ramsey interferometer or the 50:50 beam splitter in a Mach-Zehnder interferometer.

It has been demonstrated that the phase uncertainty can be enhanced beyond the standard quantum limit. The phase uncertainty $\Delta \varphi$ is inferred from the state's sensitivity to small rotation around an arbitrary axis in the $xy$-plane. When $\varphi \approx 0.015$ rad, the phase uncertainty can reduce to $-1.61_{-1.1}^{+0.98}$ dB, which is below the shot noise limit. If taking both shot noise and detection noise into account, the phase uncertainty can be improved to $-2.5_{-1.1}^{+0.98}$ dB.

\subsection{Ultracold trapped ions}

Systems of ultracold atomic ions in a Paul trap provide an excellent platform for manipulating both the internal spin and external motional degrees of freedom. Ultracold trapped ions have been proposed to explore fundamental quantum principle and implement quantum information processing~\cite{WinelandRMP, PhysRevLett.74.4091, PhysRevA.85.040302, PhysRevLett.81.5528, PhysRevLett.89.247901, Stute2012, Blinov2004, PhysRevLett.96.030404, Nature.453.1008}, quantum simulation~\cite{NatPhys.8.277, RepProgPhys.75.024401} and quantum metrology~\cite{PhysRevA.85.043604, PhysRevLett.86.5870,  PhysRevLett.89.247901, PhysRevLett.90.143602, Leibfried04062004, Roos2006}. Here, we give a brief introduction for some typical experiments of quantum metrology via ultracold trapped ions.

In 2001, Meyer et al. experimentally demonstrated that the sensitivity of rotation angle estimation with a Ramsey spectroscopy can be improved by using entangled trapped ions~\cite{PhysRevLett.86.5870}. The experiment used two $^{9}$Be$^{+}$ ions that are confined in a linear radio-frequency trap. Two hyperfine states $\left|F=1,m_F=-1\right\rangle \equiv \left|\uparrow \right\rangle$ and $\left|F=2,m_F=-2\right\rangle \equiv \left|\downarrow \right\rangle$ form the basis of an effective spin-1/2 system. $\left|\uparrow \right\rangle$ and $\left|\downarrow \right\rangle$ are coupled by two-photon Raman transitions. The detection of the ions in states $\left|\uparrow \right\rangle$ and $\left|\downarrow \right\rangle$ is done by state-sensitive fluorescence. The use of entangled states for parity measurement and Ramsey spectroscopy has been demonstrated with $\left|\Psi_P \right\rangle =\left(e^{i\phi}\left|\uparrow \uparrow \right\rangle +\left|\downarrow \downarrow \right\rangle\right)/\sqrt{2}$ and $\left|\Psi_R \right\rangle =\left(\left|\uparrow \downarrow \right\rangle +\left|\downarrow \uparrow \right\rangle\right)/\sqrt{2}$, respectively. The experimental data show the measurement sensitivity is improved beyond the standard quantum limit (SQL) and close to the Heisenberg limit (HL).

\begin{figure}[htb]
\centering
\includegraphics[width=4in]{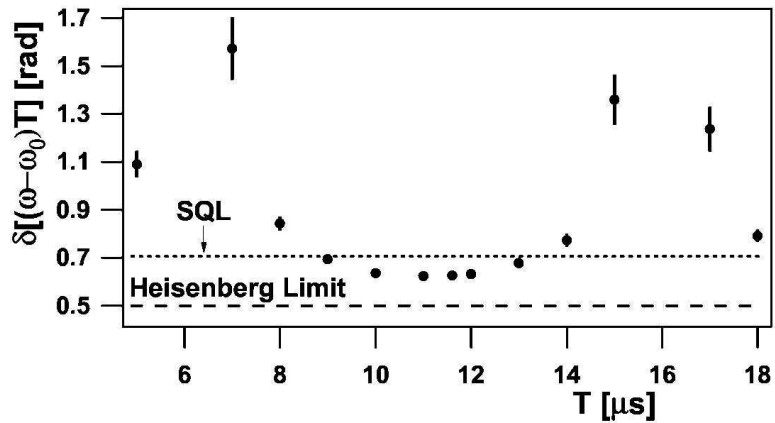}\caption{Measurement precision in a Ramsey experiment with the initial state $|\Psi_R \rangle=\left(\left|\uparrow \downarrow \right\rangle +\left|\downarrow \uparrow \right\rangle\right)/ \sqrt{2}$. The dotted line represent the SQL for two ions in a CSS. The dashed line is the Heisenberg limit. From Ref.~\citen{PhysRevLett.86.5870}.}
\label{Fig_Wineland}
\end{figure}

In 2004, Leibfried et al. demonstrated the Ramsey spectroscopy with three entangled $^{9}Be^{+}$ ions in the GHZ state. The experimental data shows that the spectroscopic sensitivity is 1.45(2) times as high as that of a perfect experiment with three independent ions, which approaches the Heisenberg limit~\cite{Leibfried04062004}.

In 2006, Roos et al. demonstrated precision spectroscopy of a pair of trapped Ca$^+$ ions in a decoherence-free subspace of specifically designed entangled states~\cite{Roos2006}. In addition to the enhancement of signal-to-noise ratio in frequency measurements, a suitably designed pair of ions enable atomic clock measurements in the presence of magnetic field noise.

There are lots of experiments about the use of trapped atomic ions in quantum metrology. More information of this field can be found in a review of Wineland and Leibfried~\cite{WinelandReview2011}.

\subsection{Cold atomic ensembles}

In addition to the realistic inter-particle interaction, quantum non-demolition (QND) has been widely used to generate quantum spin squeezing and entanglement~\cite{PhysRevLett.106.133601,12460641,Anne2010,PhysRevD.65.022002,PhysRevLett.107.080503,PhysRevLett.85.1594,
PhysRevLett.83.1319,PhysRevLett.105.193602}. It has been demonstrated that the quantum spin squeezing and entanglement for over 100 thousand cold Cs atoms can be achieved by QND measurement on the atom clock levels~\cite{12460641,Anne2010}. In the experiment, the two hyperfine states $\left|\uparrow\right\rangle \equiv |F=4, m_F=0\rangle$ and $\left|\downarrow\right\rangle \equiv |F=3, m_F=0\rangle$ of Cs atoms are referred to clock levels.

\begin{figure}[htb]
\centering
\includegraphics[width=3in]{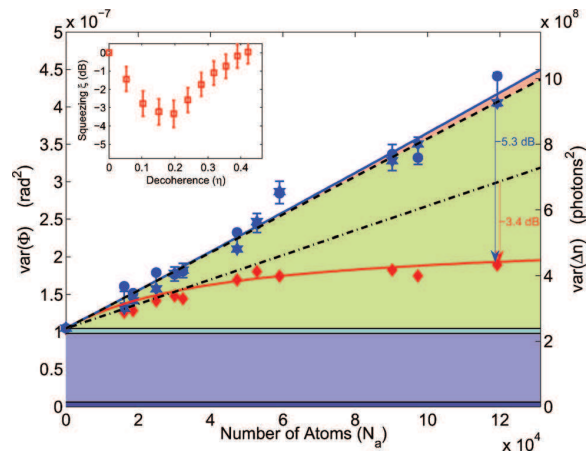}\caption{Projection noise and spin squeezing via QND. Blue points, stars: variances $\textrm{Var}(\phi_1)$, $\textrm{Var}(\phi_2)$ of $J_z$ of atoms in a CSS versus $N_A$; solid blue line: quadratic fit; dashed line: CSS projection noise; red diamonds: conditionally reduced variance of a second $J_z$ measurement predicted by the first variance; red line: reduced noise of SSS predicted from quadratic fits to projection noise data. Blue area, optical shot noise (light blue) and detector noise (dark blue); green area, projection noise. From Ref.~\citen{12460641}.}
\label{Fig_Polzik2}
\end{figure}

The experiment is implemented as follows. Initially, by using optical pumping, the Cs atoms are prepared in the clock state $\left|\downarrow\right\rangle$. To prepare the CSS, a resonant $\frac{\pi}{2}$ microwave pulse at the clock frequency is applied. Then, successive QND measurements of the population difference $N_{\uparrow}-N_{\downarrow}$ are performed by measuring the state dependent phase shift of the off-resonant probe light in a balanced homodyne configuration. After the QND measurement, all atoms are pumped into the $F=4$ level to determine the total atom number $N_A$. Two identical linear polarized beams $P_{\uparrow}$ and $P_{\downarrow}$ off-resonantly probe the transitions $|F=3\rangle$ to $|F'=4\rangle$ and $|F=4\rangle$ to $|F'=5\rangle$, respectively. Each beam gains a phase shift proportional to the number of atoms in the corresponding clock states,
\begin{eqnarray}
\phi_{\uparrow}=k_{\uparrow}N_{\uparrow},
\phi_{\downarrow}=k_{\downarrow}N_{\downarrow},\nonumber
\end{eqnarray}
where $k_{\uparrow({\downarrow})}$ are the coupling constants and the detuning $\Delta_{\uparrow(\downarrow)}$ are tuned to make $k_{\uparrow}=k_{\downarrow}=k$. The phase difference between the two arms of the Mach-Zehnder interferometer is related to the measurement of $J_z$ and the shot noise of the photons,
\begin{eqnarray}
\phi=\frac{\delta n}{n} + 2k J_z.
\end{eqnarray}
The variance of the phase difference,
\begin{eqnarray}
\textrm{Var}(\phi)=\frac{1}{n} + k^2 \textrm{Var}(\Delta N).
\end{eqnarray}
For an atomic CSS, $\textrm{Var}(\Delta N)=N_A$. Finally, use $n_1$ photons to measure $J_z$ to obtain a measurement result of $\phi_1$, then use $n_2$ photons to measure $J_z$ to obtain a measurement result of $\phi_2$ on the same atomic ensemble can create a conditionally spin squeezed atomic state. The projection noise has been reduced to $-(5.3\pm0.6)$ dB and metrologically relevant spin squeezing of $-(3.4\pm0.7)$ dB on the Cs microwave clock transition has been realized.

In addition to the above measurement-based squeezing, it has been demonstrated that spin squeezed states can be produced unconditionally by cavity feedback~\cite{PhysRevLett.104.073602}. The cavity feedback method generates spin dynamics~\cite{PhysRevA.81.021804} similar to the one-axis twisting~\cite{PhysRevA.47.5138}. By using the spin squeezed states prepared by cavity feedback method, it has been achieved a high-precision atomic clock of an ensemble of laser-cooled $^{87}$Rb atoms beyond the SQL~\cite{PhysRevLett.104.250801}. The Allan deviation spectrum indicates that the clock has a precision 2.8(3) times faster than the SQL for averaging times up to 50 s.

\section{Summary and discussion}

We have given a brief introduction on quantum metrology with cold atoms both in theory and experiment. We start from the general process of physical measurements in quantum mechanics and then discuss how to estimate an unknown parameter, which is the central goal of metrology. The estimation precision is quantified by the uncertainty, which is determined by the input state, the dynamical evolution process and the readout strategy. The uncertainty of an estimated parameter is limited by the Cram\'{e}r-Rao bound, which is related to the Fisher information. For a given input state and dynamical evolution, through optimizing over all possible measurements, there exits an ultimate precision limit determined by the quantum Fisher information.

To illustrate the general procedures of quantum metrology, we have introduced two typical quantum interferometry processes: Mach-Zehnder interferometry and Ramsey interferometry. The measurement precision of the interferometers with non-entangled states is limited by the standard quantum limit(SQL). By employing quantum entanglement, the SQL for the measurement precision can be surpassed by inputting multiparticle entangled states, such as, spin squeezed states, NOON states, entangled coherent states and twin Fock states. In realistic systems of cold atoms, the nonlinearity originated from intrinsic or laser induced atom-atom interactions can be used to generate various entangled states and then one can implement some particular precision measurements with the prepared entangled states.

Although there emerge great achievements in quantum metrology with cold atoms, to implement precision measurements with multiparticle entangled states and build practical quantum devices, there are still lots of important things need to be done. For an example, it is worthwhile to analyze the robustness against the environment effects. Therefore, it is vital to explore the effects to the measurement precision in the presence of decoherence~\cite{PhysRevA.82.045601, PhysRevA.84.043628, PhysRevA.80.053613, PhysRevLett.106.153603, PhysRevLett.111.090801, arXiv:1307.2558, arXiv:1212.3286, NewJPhys.15.073043, arXiv:1307.0470}, temperature~\cite{Sinatra2012}, and particle losses~\cite{PhysRevA.78.063828, LiYun2009, PhysRevA.88.013838, PhysRevA.83.063836, PhysRevA.79.023812, PhysRevLett.108.233602}.

Besides to the metrology schemes using entanglement as a resource for beating the SQL, there are some alternative entanglement-free schemes beating the SQL. By replacing entangled input states with multiple applications of the phase shift on unentangled single-photon states, the Heisenberg-limited phase estimation has been demonstrated~\cite{Nature.450.393}. By coupling the quantum resources to a common environment that can be measured at least in part, it has been shown that the Heisenberg-limited measurements can be achieved~\cite{NatComms.2.223}.

Beyond measuring single parameters in most metrology schemes, it is also possible to estimate simultaneously multiple parameters in some particular metrology schemes. Based upon a discretized phase imaging model, which is an interferometer of $(d+1)$ modes, $d$ independent phases are possible to be estimated simultaneously~\cite{PhysRevLett.111.070403}. The theoretical analysis shows that, (i) the quantum strategies for both independent and simultaneous parameter estimation follow the Heisenberg scaling in the total number of photons for the total variance, and (ii) simultaneous quantum phase estimation improves the precision linearly with the number of phases (i.e. scaling as $\thicksim 1/d$). Based upon a three-dimensional waveguide, a three-arm interferometer has been proposed to achieve the simultaneous two-parameter estimation~\cite{screp.2.862}. Based upon a two-phase spin rotation, the joint and sequential quantum estimations of the two phases have been studied and the results show the joint estimation method gives a better sensitivity~\cite{QMetro.1.12}.

Lastly, we would like to point out that there are also some metrology schemes realizable with cold atoms alternative to interferometry. By calculating the quantum Fisher information matrix for quantum gases, the sensitivity of measuring the temperature and the chemical potentials of quantum gases has been investigated~\cite{arXiv:1308.2735}. The analysis shows that the SQL can be surpassed by using bosonic gases, but not for fermionic gases. The experimental realization of thermometry with Bose condensed atoms has been reported by MIT~\cite{Science.301.1513}. Based upon detection of Larmor spin precession of optically pumped atoms in a magnetic field, it is possible to detect weak magnetic field in the radio-frequency range~\cite{Nature.422.596, PhysRevLett.95.063004, arXiv:1207.2842}. For an atomic magnetometer based upon a spinor BEC with a off-resonant optical field, Steinke et al. analyzes how its sensitivity depends on the dispersive interaction between the spinor BEC and the off-resonant optical field~\cite{arXiv:1211.287}. More recently, the detection of a weak alternate-current magnetic field has been demonstrated by applying spin-echo techniques to a spin-2 atomic BEC~\cite{arXiv:1306.1011}.

\section*{Acknowledgement}

We thank Zbigniew Ficek, Roberto Floreanini, Ugo Marzolino, Gheorghe Sorin Paraoanu, Christopher Gerry, and Bruno Julia-Diaz for their useful comments and suggestion. This work is supported by the NBRPC under Grant No. 2012CB821305, the NNSFC under Grants No. 11075223 and No. 11374375, the NCETPC under Grant No. NCET-10-0850, and the Ph.D. Programs Foundation of Ministry of Education of China under Grant No. 20120171110022.

\bibliographystyle{ws-rv-van}

\printindex                         

\end{document}